\documentclass[10pt]{article} % For LaTeX2e

\usepackage[accepted]{rlj} % Should be uncommented for the camera-ready
\usepackage{amssymb}            % Defines common symbols like \mathbb R
\usepackage{mathtools}          % Extends amsmath, providing common math tools
\usepackage{mathrsfs}           % Enables \mathscr, which can work in cases that \mathcal does not
\usepackage{graphicx}           % For including images
\usepackage{subcaption}         % Allows for the use of subfigures and subcaptions
\usepackage[space]{grffile}     % For spaces in image names
\usepackage{url}                % For displaying URLs
\usepackage{lipsum}             % For placeholder text

\title{TransAM: Transformer-Based Agent Modeling for Multi-Agent Systems via Local Trajectory Encoding}

\setrunningtitle{Transformer-Based Agent Modeling for Multi-Agent Systems}

\author{Conor Wallace, Umer Siddique, Yongcan Cao}

\emails{conor.wallace@my.utsa.edu \\ muhammadumer.siddique@my.utsa.edu, yongcan.cao@utsa.edu}

\affiliations{
\textbf{Department of Electrical and Computer Engineering, \\University of Texas at San Antonio}\\
}

\contribution{
    We eliminate the need for access to other agents' trajectories at inference time by learning a latent policy representation derived solely from the local trajectory of the controlled agent.
    }
    {
    It is common for agent modeling methods to assume access to other agent information at execution time \citep{dron, grover, tao}.
    }

\contribution{
    By treating the local trajectory of the controlled agent as a temporal sequence, we use a transformer to model long-range dependencies and identify key moments that characterize interactions with other agents. This is in contrast to previous methods that rely on MLPs or RNNs without attention over extended time horizons.
    }
    {
    Other methods typically construct either an MLP-based agent model \citep{dron}, or a recurrent agent model \citep{liam}, which do not take into account the full context of the agent's trajectory throughout the episode.
    }

\contribution{
    To address the data demands of transformers, we train the agent model and the controlled agent's policy jointly in an online setting, ensuring access to a diverse dataset for enhanced performance.
    }
    {
    Other promising transformer-based agent modeling approaches, such as \citep{tao} are based on an offline reinforcement learning setting wherein a pretraining phase is used to learn an initial prior for the task. In contrast, we aim to train the agent model and the policy jointly from scratch.
    }

 % Your keywords

\summary{Agent modeling is a critical component in developing effective policies within multi-agent systems, as it enables agents to form beliefs about the behaviors, intentions, and competencies of others. Many existing approaches assume access to other agents' episodic trajectories, a condition often unrealistic in real-world applications. Consequently, a practical agent modeling approach must learn a robust representation of the policies of the other agents based only on the local trajectory of the controlled agent. In this paper, we propose \texttt{TransAM}, a novel transformer-based agent modeling approach to encode local trajectories into an embedding space that effectively captures the policies of other agents. We evaluate the performance of the proposed method in cooperative, competitive, and mixed multi-agent environments. Extensive experimental results demonstrate that our approach generates strong policy representations, improves agent modeling, and leads to higher episodic returns.
}

\begin{document}

% \makeCover  % Create the cover page
\maketitle  % Make the title section

\begin{abstract}
Agent modeling is a critical component in developing effective policies within multi-agent systems, as it enables agents to form beliefs about the behaviors, intentions, and competencies of others. Many existing approaches assume access to other agents' episodic trajectories, a condition often unrealistic in real-world applications. Consequently, a practical agent modeling approach must learn a robust representation of the policies of the other agents based only on the local trajectory of the controlled agent. In this paper, we propose \texttt{TransAM}, a novel transformer-based agent modeling approach to encode local trajectories into an embedding space that effectively captures the policies of other agents. We evaluate the performance of the proposed method in cooperative, competitive, and mixed multi-agent environments. Extensive experimental results demonstrate that our approach generates strong policy representations, improves agent modeling, and leads to higher episodic returns.
\end{abstract}

%%%%%%%%%%%%%%%%%%%%%%%%%%%%%%%%%%%%%%%%%%%%%%%%%%%%%%%%%%%%%%%%
%% Section: Main Text
%%%%%%%%%%%%%%%%%%%%%%%%%%%%%%%%%%%%%%%%%%%%%%%%%%%%%%%%%%%%%%%%

\section{Introduction}
\label{sec:intro}

Recent advances in multi-agent systems have led to significant progress in domains such as games~\citep{nowe2012game}, traffic control~\citep{wiering2000multi}, and autonomous driving~\citep{cao2012overview}. A key challenge in these systems is that the actions of all agents influence the overall system's transitions. Therefore, effectively reasoning about the optimal actions requires modeling the behavior of other agents. This process, known as agent modeling, focuses on inferring concealed information about other agents to inform the policy of a controlled agent. In this work, we explore the role of agent modeling in multi-agent systems and its impact on decision-making strategies.

A primary challenge in agent modeling arises from the need to design agents that can adapt to various agent policies using only the information available during execution. This challenge becomes particularly difficult in scenarios where no direct information about the other agents is accessible, requiring the agent to infer others' behaviors based solely on its own local information. Moreover, since agent policies may appear indistinguishable on the basis of a single transition, it is essential to consider the temporal context for disambiguation. Therefore, an effective agent modeling approach must learn robust representations of agent policies while accounting for their temporal dynamics and long-term effects.

Although recent advances in deep learning have led to various approaches for agent modeling~\citep{dron, grover, liam, tao}, existing methods often face two key limitations: (1) reliance on access to agent trajectories and (2) inadequate use of the sequence of actions of the controlled agent as a valuable source of information. Inspired by the success of decision transformers~\citep{decision_transformer} and their multi-agent variants~\citep{mat}, we propose reframing agent modeling as a sequence modeling task using a transformer architecture.

Transformers have recently been applied in reinforcement learning (RL) and demonstrated remarkable success, from feature extraction to end-to-end policy learning~\citep{transformer_rl_survey}. Building on this, we propose a transformer-based agent modeling approach that encodes the controlled agent's local trajectory into an embedding space that captures the influence of other agent policies. The model is trained to reconstruct the other agents' trajectories using only the local embedding, enabling the controlled agent to model others without requiring access to their trajectories at execution. This allows the RL policy to condition its decisions solely on the local trajectory embeddings.

Our contributions are as follows.

\begin{enumerate}[leftmargin=*, label=\arabic*., before=\setlength{\itemindent}{0pt}]
    \item \textbf{Agent Modeling from Local Information:} We eliminate the need for access to other agents' trajectories at inference time by learning a latent policy representation derived solely from the local trajectory of the controlled agent.
    \item \textbf{Local Trajectory as a Sequence Modeling Task:} By treating the local trajectory of the controlled agent as a temporal sequence, we use a transformer to model long-range dependencies and identify key moments that characterize interactions with other agents.
    \item \textbf{Online Joint Training of Agent Model and Policy:} Unlike prior agent modeling methods that pretrain a transformer encoder, we train the agent model and the controlled agent’s policy jointly in an online setting.
\end{enumerate}

We evaluate the proposed approach on cooperative, competitive, and mixed cooperative-competitive multi-agent RL tasks. Our results demonstrate that the proposed method outperforms baseline approaches in agent modeling accuracy, provides robust agent policy representation, and achieves superior episodic returns.

\section{Related Work}
\label{sec:related}

\subsection{Agent Modeling}
When operating in a decentralized multi-agent system, it is important to incorporate information about other agents to determine the best response to a given state. In conventional centralized training with decentralized execution (CTDE) approaches, such as MADDPG \citep{maddpg} and MAPPO \citep{mappo}, a centralized critic is trained using the joint observations of all agents, and this information is implicitly distilled into the actor policy. Agent modeling is an alternative approach that explicitly learns to model concealed agent information. There is a large body of work on agent modeling in multi-agent settings \citep{opponent_modeling_survey}. \citet{dron} focused on competitive settings and learned to predict opponent Q values and opponent actions given opponent observations. \citet{som} introduced a model that learns to infer the opponent's goal using itself. \citet{grover} implemented a general purpose encoder-decoder architecture using imitation learning and a contrastive triplet loss to both learn to accurately reconstruct agent policies and correctly identify the agent policy within the embedding space. Building on the work of \citet{grover}, \citet{liam} also used an encoder-decoder architecture to reconstruct agent policies. However, they model this reconstruction using the controlled agent's local trajectory only. \citet{fastap} introduced an approach that adapts to changing policies, similar to our problem setting. However, agents in this work can change policies within an episode, so the model must learn to quickly adapt. \citet{ctcat} studied ad hoc teamwork in which an agent must learn to cooperate with other agents who may switch to different goal-oriented policies. In this work, the agent learns both to identify the type of policy of its teammates and to generalize the types of policies to unseen sets of teammates. Finally, \citet{clam} learned an agent policy representation directly from the controlled agent's local observations using contrastive learning.

\subsection{Transformers in RL}

Transformers were originally intended as replacements for RNNs in machine translation language modeling tasks \citep{transformer}. However, they have been applied to seemingly every subfield of machine learning, including computer vision \citet{vit} and more recently for reinforcement learning \citep{transformer_rl_survey}. The original transformer model consists of an encoder that maps an input sequence to a latent space and a decoder that generates an output sequence conditioned on the input sequence and the latent embeddings of the input sequence. Reinforcement learning problems have incorporated both parts of the transformer model to pose the problem in different terms. \citet{gtrxl} used a modified encoder architecture as a replacement for RNNs in RL policies. Alternatively, \citet{decision_transformer} proposed offline RL as a generative sequence modeling task using a GPT-style decoder architecture \citep{gpt}. More recently, multi-agent reinforcement learning has been reimagined as a sequence-to-sequence task \citep{mat} where the model maps input sequences of observations to output sequences of actions. Similarly to our problem setting, \citet{tao} introduced a transformer architecture to learn opponent policy representations from offline datasets. In this paper, we are interested in learning latent representations of the other agents' policies as a function of the controlled agent's local trajectory.
\section{Background}
\label{sec:background}

\subsection{Partially Observable Stochastic Games}
Partially observable stochastic games (POSGs) \citep{posg} are a common formulation for multi-agent settings. They are described by a set of agents $i \in \{0, \ldots, N\}$ and a finite set of states $s \in \mathcal{S}$. For each agent $i$, there is a finite action space $\mathcal{A}^{i}$ where $\mathcal{A} = \mathcal{A}^{0} \times \ldots \times \mathcal{A}^N$ represents the joint action space of all agents. Similarly, for each agent $i$, there is a finite observation space $\mathcal{O}^{i}$, where $\mathcal{O} = \mathcal{O}^{0} \times \ldots \times \mathcal{O}^{N}$ is the joint observation space of all agents. In addition to the observation space, an agent has an observation function $O^{i}$: $\mathcal{A} \times \mathcal{S} \times \mathcal{O}^{i} \rightarrow [0, 1]$ given by \ref{eq:observation_fn}
\begin{equation}
    \label{eq:observation_fn}
    \forall a \in \mathcal{A}, \forall s \in \mathcal{S}: \sum_{o^{i} \in \mathcal{O}^{i}}O(a, s, o^{i}) = 1.
\end{equation}
In addition to the action and observation spaces, each agent has a reward function $\mathcal{R}^{i}:\mathcal{S} \times \mathcal{A} \times \mathcal{S} \rightarrow \mathbb{R}$. Finally, similar to the observation function, the game has a state transition probability function $\mathcal{P}:\mathcal{S} \times \mathcal{A} \times \mathcal{S} \rightarrow [0, 1]$ given by \ref{eq:transition_fn}
\begin{equation}
    \label{eq:transition_fn}
    \forall a \in \mathcal{A}, \forall s \in \mathcal{S}: \sum_{s' \in \mathcal{S}}P(s, a, s') = 1,
\end{equation}
where $s'$ is the next state as a result of taking the joint action $a$ in the previous state $s$.

Agent $i$ selects an action $a^{i} \in \mathcal{A}^i$ given an observation $o^{i} \in \mathcal{O}^{i}$ according to a policy $\pi^{i}(a^{i}|o^{i})$, which is a probability distribution over the set of actions $\mathcal{A}^{i}$. The goal of an agent is to learn a policy $\pi$ such that the expected cumulative reward, or the agent's return, is maximized:
\begin{equation}
\label{eq:policy_obj}
\max_\pi \mathbb{E} \left[ \sum_{t=1}^{L} \gamma^{t} r_{t+1} \mid\pi\right]
\end{equation}
where $L$ is the length of the episode and $\gamma \in [0, 1)$ is the discount factor. The action value function $Q^{\pi^{i}}(s, a^{i})$ for agent $i$ defines the expectation of the return given the state $s$ when taking action $a^{i}$ following policy $\pi^{i}$. Similarly, the value function $V^{\pi^{i}}(s)$ describes the value of being in state $s$ for agent $i$ following policy $\pi^{i}$. In actor-critic methods, such as A2C \citep{a2c}, the actor $\pi^{i}$ and the critic $V^{\pi^{i}}(s)$ are used to calculate the advantage function $A^{\pi^{i}}(s, a^{i}) = Q^{\pi{^{i}}}(s, a^{i}) - V^{\pi^{i}}(s)$.

\subsection{Transformers}
% make sure that the notation is specific for each item we are referencing, e.v. Q^pi vs. query Q, etc.
Transformers consist of an encoder and a decoder and can use either the encoder, the decoder, or both depending on the applications. Generalizing, encoder-decoder models are used for machine translation tasks \citep{t5}. Decoder-only models are useful for generative sequence tasks \citep{gpt}. Encoder-only models are good for sequence understanding tasks \citep{bert}. We make use of an encoder-only model for our problem, and hence will focus on this portion of the model. The encoder takes as input a sequence of embedding tokens $\{T_{t}, \ldots, T_{t+K}\}$ with context length $K$ and transforms them into representation embedding vectors $\{E_{t}, \ldots, E_{t+K}\}$. The model is composed of several layers of transformer blocks. Each block contains a multi-head self-attention layer and a feed-forward layer, connected by a residual connection with layer normalization at the output of the block. The self-attention function below uses three linear layers to map the input sequence of the $i^{th}$ block into query $\mathcal{Q}_{i}$, key $\mathcal{K}_{i}$, and value $\mathcal{V}_{i}$ matrices which are used to create the output as follows
\begin{equation}
    \label{eq:attention}
    \mathcal{Z}_{i} = \text{softmax} \left( \frac{\mathcal{Q}_{i}\mathcal{K}_{i}^{T}}{\sqrt{d_{k}}} \right) \mathcal{V}_{i},
\end{equation}
where $d_{k}$ is the dimension of the input token vectors. By combining the input tokens into sequence matrices $\mathcal{Q}$, $\mathcal{K}$, and $\mathcal{V}$ the self-attention function attends to the whole sequence, allowing the model to extract relevant information throughout the sequence.

\begin{figure*}[t!]
\centering
\includegraphics[width=0.99\textwidth]{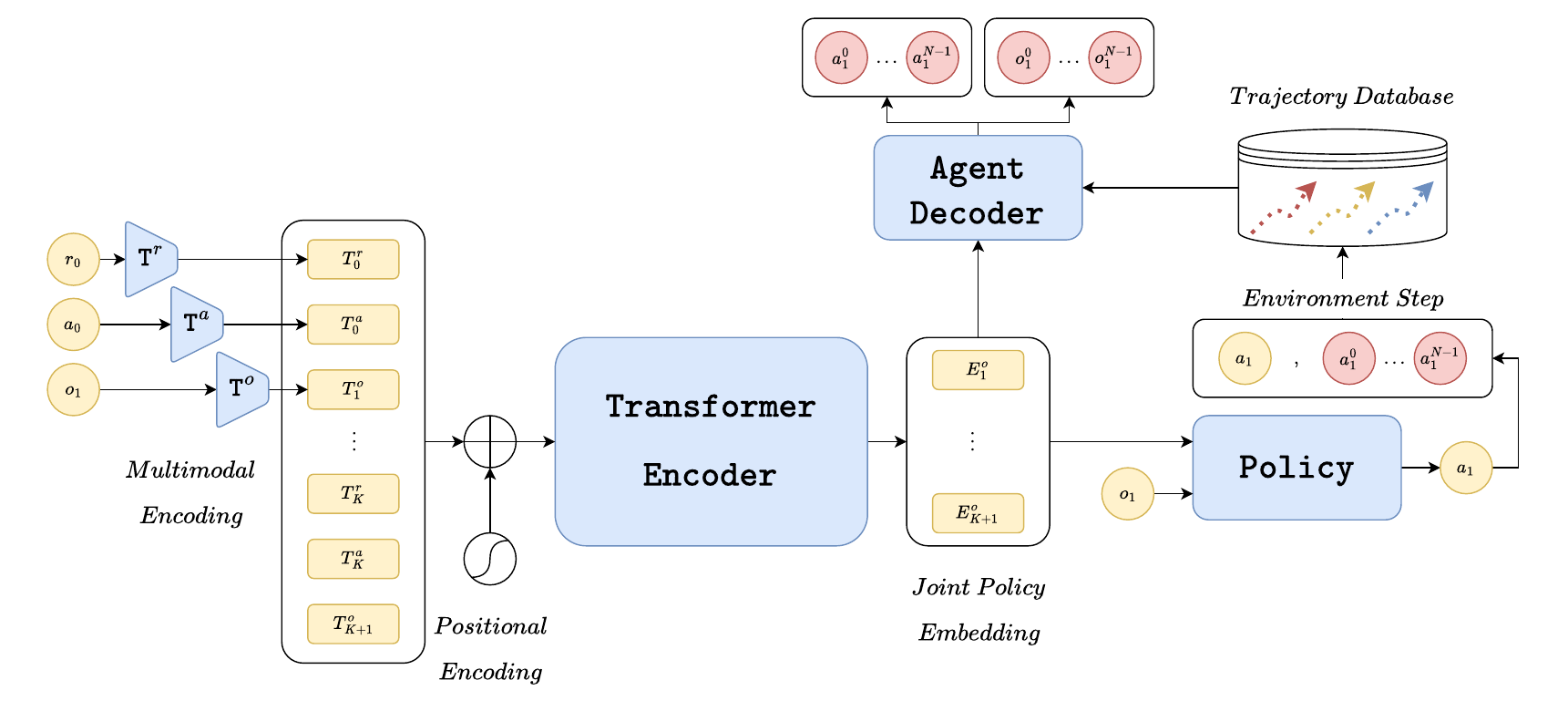}
\caption{\textbf{\texttt{TransAM} architecture.} We embed the controlled agent's previous reward, previous action, and current observation into embedding tokens, $T_{t}^{(r, a, o)}$, and transform them into an output sequence of embedding vectors, $E_{t}^{(r, a, o)}$. The embedding vectors are used to both condition the controlled agent's policy and reconstruct the other agents' trajectories as a function of the local trajectory only.}
\label{fig:pipeline}
\end{figure*}

\subsection{Problem Formulation}

We consider a modified POSG with one learning agent under our control and a set of agents to interact with, which can utilize one of several fixed policies. To be specific, we assume that each individual agent $i$ adopts a policy $\pi^{i}$, whose collection forms the joint agent policy $\pi^{-1}$. We consider the set of $M$ fixed, pre-trained joint policies $\Pi = \{\pi^{-1, m}| m = 1, \ldots , M\}$, composed of agents trained via heuristic and reinforcement learning strategies. For clarity, we refer to the controlled agent without a superscript and to all other agents with superscript -$1$. Thus, the controlled agent has an action space $\mathcal{A}$ and an observation space $\mathcal{O}$, while other agents have joint spaces $\mathcal{A}^{-1}$ and $\mathcal{O}^{-1}$. Our goal is to learn a policy $\pi_{\theta}$ parameterized by $\theta$ that maximizes the expected return averaged over all joint policies $\Pi$ by optimizing Equation \ref{eq:modified_policy_obj}:

\begin{equation}
\label{eq:modified_policy_obj}
\arg \max_{\theta} \mathbb{E}_{\pi_{\theta}, \pi^{-1, m} \sim \mathcal{U}(\Pi)} \left[ \sum_{t=1}^{L} \gamma^{t} r_{t+1} \right],
\end{equation}
where $\pi^{-1,m}$ is uniformly sampled from $\Pi$ at the beginning of each episode. The agent policy type $m$ is concealed from the controlled agent throughout the episode. This occluded information can either be incorporated into the policy implicitly by simply attempting to maximize the average return for all agent policies, or it can be modeled explicitly and used to condition the policy on which policy $m$ is currently being modeled. In this work, we focus on the latter and introduce a transformer-based approach to modeling such agent policies.
\section{Method}
\label{sec:method}

\subsection{TransAM}

% I think we can use either view or design instead of format?
We format agent modeling as a sequence modeling task through the lens of episodic trajectories. Consider the tuple $(r_{t-1}, a_{t-1}, o_{t})$ where $r_{t-1} \sim \mathcal{R}$ is the previous reward, $a_{t-1} \sim \mathcal{A}$ is the previous action, and $o_t \sim \mathcal{O}$ is the current observation of the controlled agent. The local episodic trajectory of the agent can be viewed as a sequence of these tuples $\mathcal{T} = (r_{0}, a_{0}, o_{1}, \ldots, r_{L-1}, a_{L-1}, o_{L})$. Similarly, other agent trajectories are represented as $\mathcal{T}^{i, m} = (r_{0}^{i, m}, a_{0}^{i, m}, o_{1}^{i, m}, \ldots, r_{L-1}^{i, m}, a_{L-1}^{i, m}, o_{L}^{i, m})$. Our goal here is to learn a representation of the joint agent policy $\pi^{-1, m}$ such that this representation can be used as an inductive bias for the controlled agent policy. Inspired by the recent success of transformers in such problems, we built a transformer encoder model, which we refer to as Transformer-based Agent Modeling (\texttt{TransAM}), to encode these sequences into a compact representation. Our proposed architecture can be seen in Figure \ref{fig:pipeline}.

We learn a linear mapping from $r_{t}$, $a_{t}$, $o_{t+1}$ to token embeddings $T^{r}_{t}$, $T^{a}_{t}$, and $T^{o}_{t+1}$, respectively. Considering the three modalities, we use a context window of $3K$ tokens as a subset of the agent's local trajectory $\mathcal{T}_{t+K} = (T^{r}_{t-1}, T^{a}_{t-1}, T^{o}_{t}, \ldots, T^{r}_{t+K-1}, T^{a}_{t+K-1}, T^{o}_{t+K})$. Using the encoder, we encode this token sequence into a representation embedding sequence $\mathcal{E}_{t+K} = (E^{r}_{t-1}, E^{a}_{t-1}, E^{o}_{t}, \ldots, E^{r}_{t+K-1}, E^{a}_{t+K-1}, E^{o}_{t+K})$. We use only observation embeddings for downstream tasks, as reward/action embeddings offer marginal empirical benefit. This embedding vector $E^{o}_{t+K}$, in addition to observation $o_{t+K}$, is used to condition the policy $\pi_{\theta}(a_{t+K} | o_{t+K}, E^{o}_{t+K})$. We posit that this incorporation of information is necessary for the agent policy to accurately determine the best response to the current joint agent policy.

% Opponent Trajectory Reconstruction Head
\textbf{Generative Loss} To learn an informative representation of the joint agent policy, we introduce an agent trajectory reconstruction head. It decodes the embedding vector $E^{o}_{t}$ into the joint observations $o_{t}^{-1, m} = (o_{t}^{0, m}, \ldots, o_{t}^{N-1, m})$ and actions $(a_{t}^{0, m}, \ldots, a_{t}^{N-1, m})$ of the other agents. We use the mean squared error loss, $\mathcal{L}_{MSE}$, to learn the observations of the agent and the mean cross-entropy loss $\mathcal{L}_{CE}$ for all actions of the agents $N-1$. In total, the agent modeling loss is given by Equation (\ref{eq:am})
\begin{equation}
\label{eq:am}
    \mathcal{L}_{AM} = \mathcal{L}_{MSE}(\hat{o}_{t}^{-1, m}, o_{t}^{-1, m}) + \frac{1}{N-1} \sum_{i=0}^{N-1} \mathcal{L}_{CE} (\hat{a}_{t}^{i, m}, a_{t}^{i, m}),
\end{equation}
where $\hat{o}_{t}^{-1, m}$ is the predicted joint agent observation and $\hat{a}_{t}^{i, m}$ is the action for agent $i$. The reconstruction head is only used during training to learn the representation $E^{o}_{t}$. During execution, we only use the encoder, which does not need access to the occluded information of other agents.

\subsection{Policy Training}
The goal of the controlled agent is to learn a policy that adapts to different joint agent policies $\pi^{-1,m}$. We train \texttt{TransAM} such that the embedding vector $E^{o}_{t}$ is a good proxy for the true other agent information. By incorporating this vector into the controlled agent policy, it allows the policy to better adapt to varying joint agent policies. 
% To do this, we concatenate the agent observation, $o_{t}$, with the embedding vector, $E^{o}_{t}$, to form the input vector $x_{t}$. 
From here, any RL algorithm can be used to learn an optimal policy $\pi$ conditioned on $o_{t}$ and $E^{o}_{t}$. In this paper, we use the advantage actor-critic (A2C) algorithm \citep{a2c}. Thus, the RL objective is given by Equation (\ref{eq:a2c})
\begin{equation}
\label{eq:a2c}
\begin{split}
    \mathcal{L}_{A2C} = &\mathbb{E}_{(o_{t}, a_{t}, o_{t+1}, r_{t+1}) \sim B} [  \frac{1}{2} \left ( r_{t+1} + V_{\phi}(o_{t+1}, E^{o}_{t+1}) - V_{\phi}(o_{t}, E^{o}_{t}) \right )^{2} \\
    & - A^{\pi}(o_{t}, a_{t})\log \pi_{\theta}(a_{t} | o_{t}, E^{o}_{t}) - \beta H(\pi_{\theta}(a_{t} | o_{t}, E^{o}_{t})) ],
\end{split}
\end{equation}
where $B$ is a batch of transitions, $\pi_{\theta}$ is the policy parameterized by $\theta$, $V_{\phi}$ is the value function parameterized by $\phi$, $A^{\pi}$ is the advantage function under policy $\pi$, and $H$ is the entropy function weighted by the entropy coefficient $\beta$. We optimize Equations \eqref{eq:am} and \eqref{eq:a2c} jointly, sampling the set of other agent policies per episode.
\section{Experiments}
\label{sec:experiments}

\begin{figure}
    \centering
    \begin{subfigure}[b]{0.24\textwidth}
        \centering
        \fbox{\includegraphics[height=3cm]{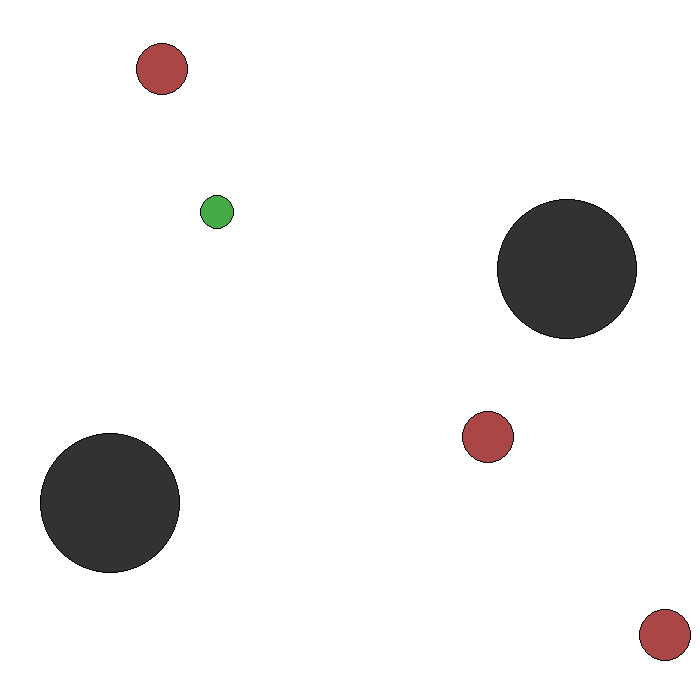}}
    \end{subfigure}
    \begin{subfigure}[b]{0.24\textwidth}
        \centering
        \fbox{\includegraphics[height=3cm]{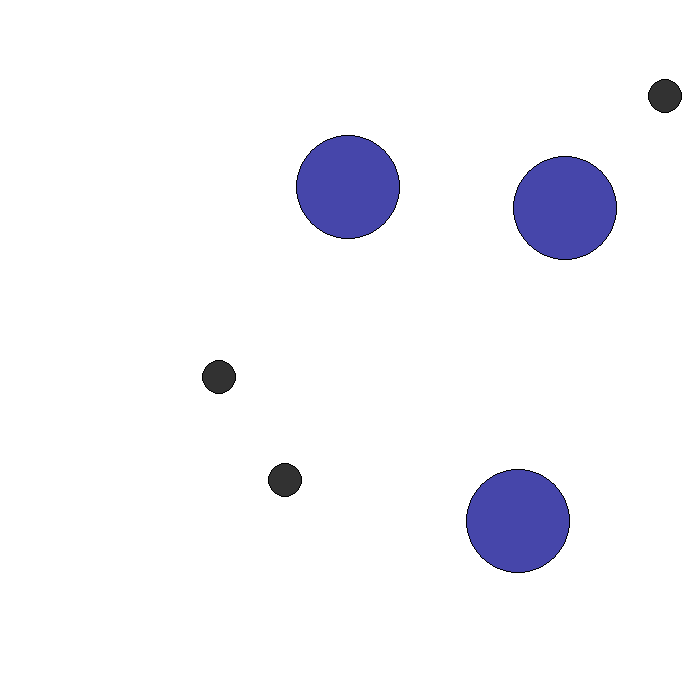}}
    \end{subfigure}
    \begin{subfigure}[b]{0.24\textwidth}
        \centering
        \fbox{\includegraphics[width=3cm, height=3cm]{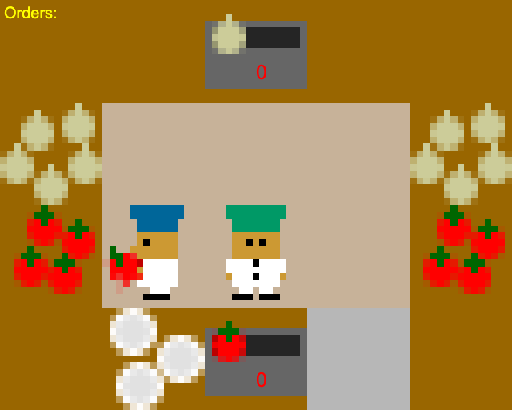}}
    \end{subfigure}
    \begin{subfigure}[b]{0.24\textwidth}
        \centering
        \fbox{\centerline{\includegraphics[width=3cm, height=3cm]{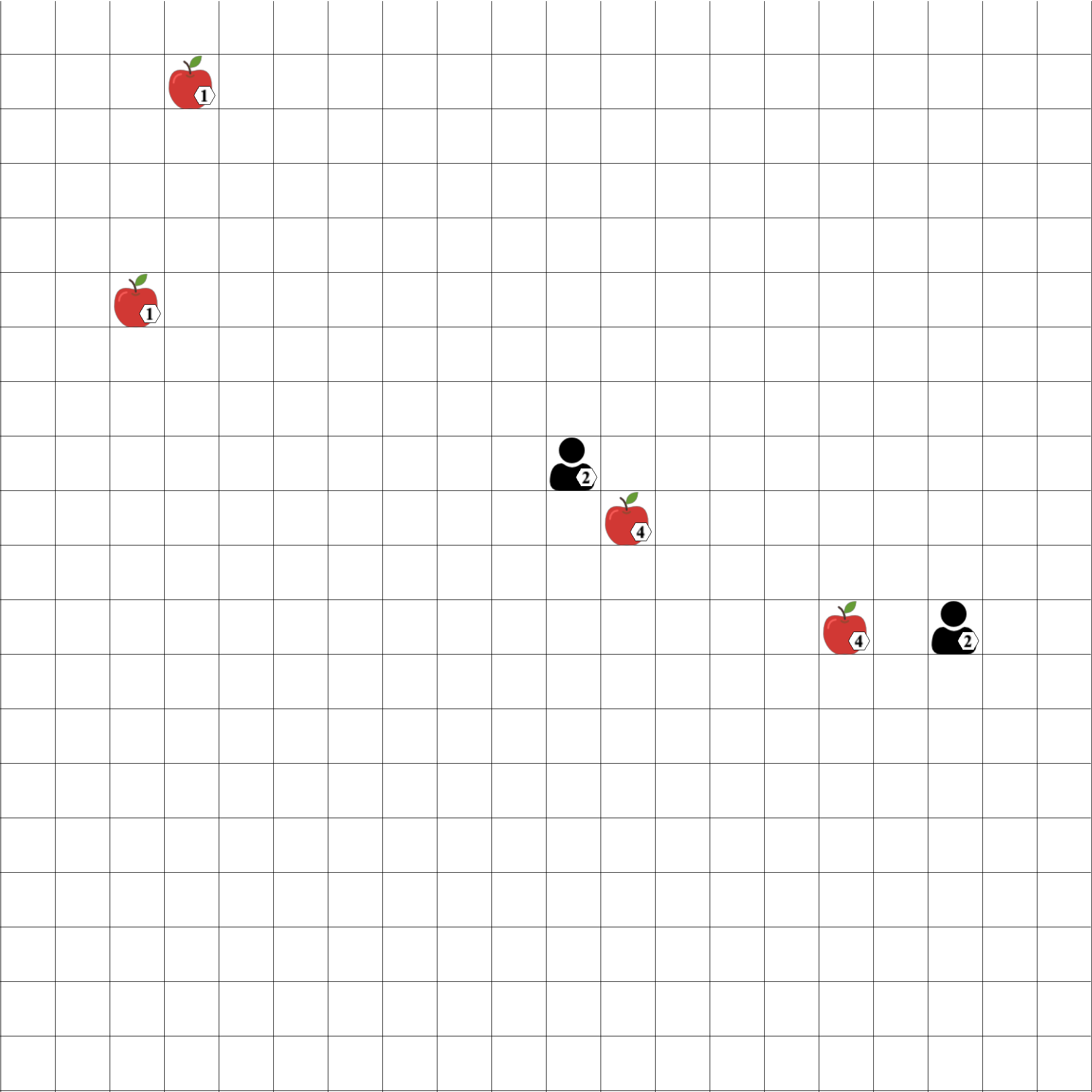}}}
    \end{subfigure}
    \caption{\textbf{Experimental environments.} We use four environments (a) Predator-Prey, a competitive pursuit environment (b) Cooperative Navigation, a cooperative navigation environment (c) Overcooked, a cooperative cooking environment (d) Level-Based Foraging a mixed resource allocation environment.}
    \label{fig:environments}
\end{figure}

\subsection{Experimental Setup}
To validate the effectiveness of our proposed approach, we performed experiments in a variety of settings, including competitive, cooperative, and mixed environments. Specifically, we used Multi-Agent Particle Environments (MPEs) from \citep{mpe} that contain competitive and cooperative scenarios, the cooperative Overcooked environment \citep{overcooked}, and the mixed level-based foraging environment \citep{lbf}. Each experiment presents a unique scenario where cooperativeness, competitiveness, or a mixture of both plays a vital role and must be modeled appropriately. Through rigorous analysis, we assessed the performance of our approach in terms of modeling agent behavior and solving the final task. In all of our experiments, we relied on the Advantage Actor-Critic (A2C) algorithm \citep{a2c} and used one LSTM layer \citep{lstm} and one linear layer, both with a hidden dimension of $128$. Furthermore, we used a transformer encoder that is made up of four transformer blocks with four attention heads and a hidden dimension of $128$. We trained the controlled agent policy for 10 million time steps and performed evaluations every 100 episodes. To ensure the reproducibility of the results, we performed five different training runs with different random seeds and plotted the average of the results to provide reliable evidence of our approach's performance.

% \begin{figure*}[t!]
% \begin{tabular}{cc}
%   \begin{subfigure}[b]{0.48\textwidth}
%     \includegraphics[clip, trim=0.4cm 0.4cm 0.4cm 0.4cm, width=\textwidth]{images/comm_return.pdf}
%     \caption{Comm returns.}
%     \label{fig:return1c}
%   \end{subfigure}
%   \begin{subfigure}[b]{0.48\textwidth}
%     \includegraphics[clip, trim=0.4cm 0.4cm 0.4cm 0.4cm, width=\textwidth]{images/push_return.pdf}
%     \caption{Push returns.}
%     \label{fig:return1c}
%   \end{subfigure} \\
%   \begin{subfigure}[b]{0.48\textwidth}
%     \includegraphics[clip, trim=0.4cm 0.4cm 0.4cm 0.4cm, width=\textwidth]{images/spread_return.pdf}
%     \caption{Spread returns.}
%     \label{fig:return1a}
%   \end{subfigure} &
%   \begin{subfigure}[b]{0.48\textwidth}
%     \includegraphics[clip, trim=0.4cm 0.4cm 0.4cm 0.4cm, width=\textwidth]{images/tag_return.pdf}
%     \caption{Tag returns.}
%     \label{fig:return1b}
%   \end{subfigure} \\
% \end{tabular}
% \caption{Average episodic returns and 95\% confidence intervals for the two experimental scenarios across ten random seeds.}
% \label{fig:returns}
% \end{figure*}

\begin{figure}
    \centering
    \begin{subfigure}[b]{.245\textwidth}
        \includegraphics[width=\linewidth]{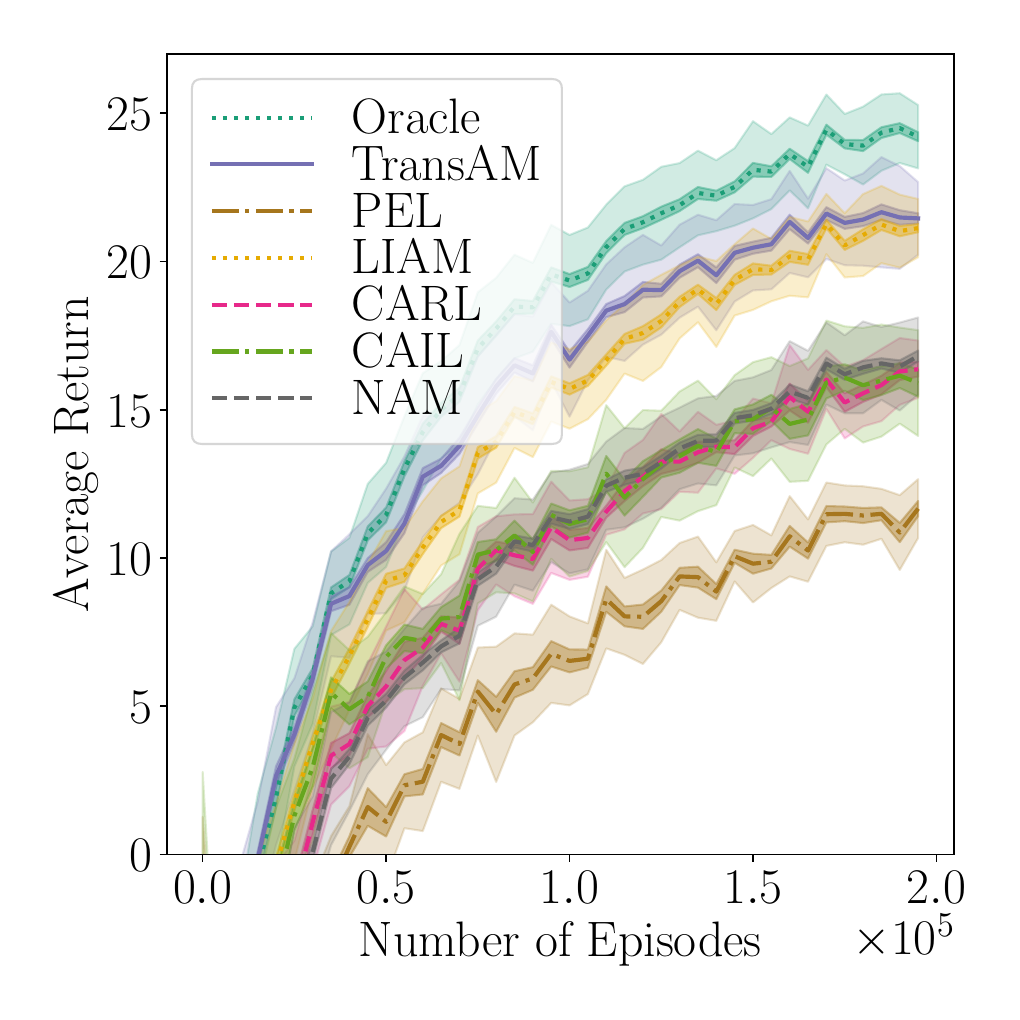}
        % \caption{Comm returns.}
        \label{fig:sub1}
    \end{subfigure}
    \begin{subfigure}[b]{.245\textwidth}
        \includegraphics[width=\linewidth]{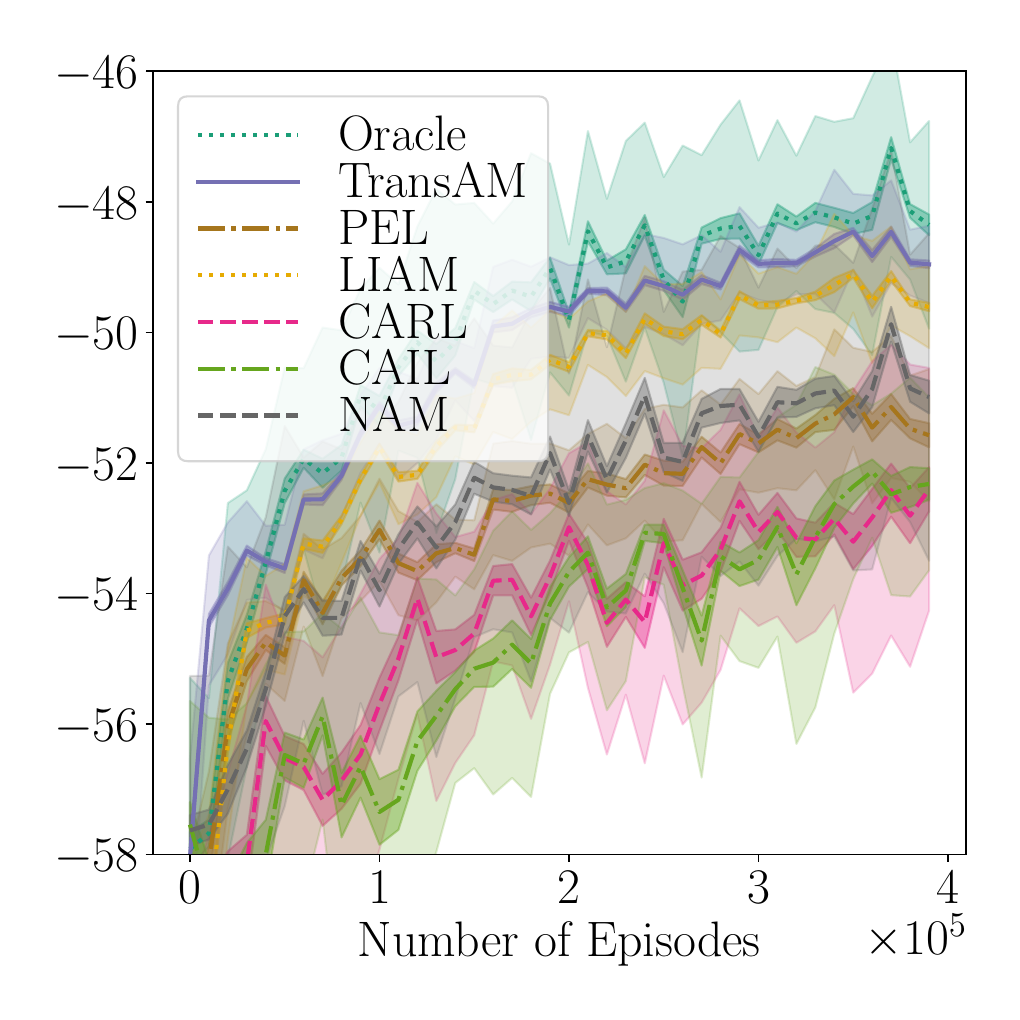}
        % \caption{Tag returns.}
        \label{fig:sub2}
    \end{subfigure}
    \begin{subfigure}[b]{.245\textwidth}
        \includegraphics[width=\linewidth]{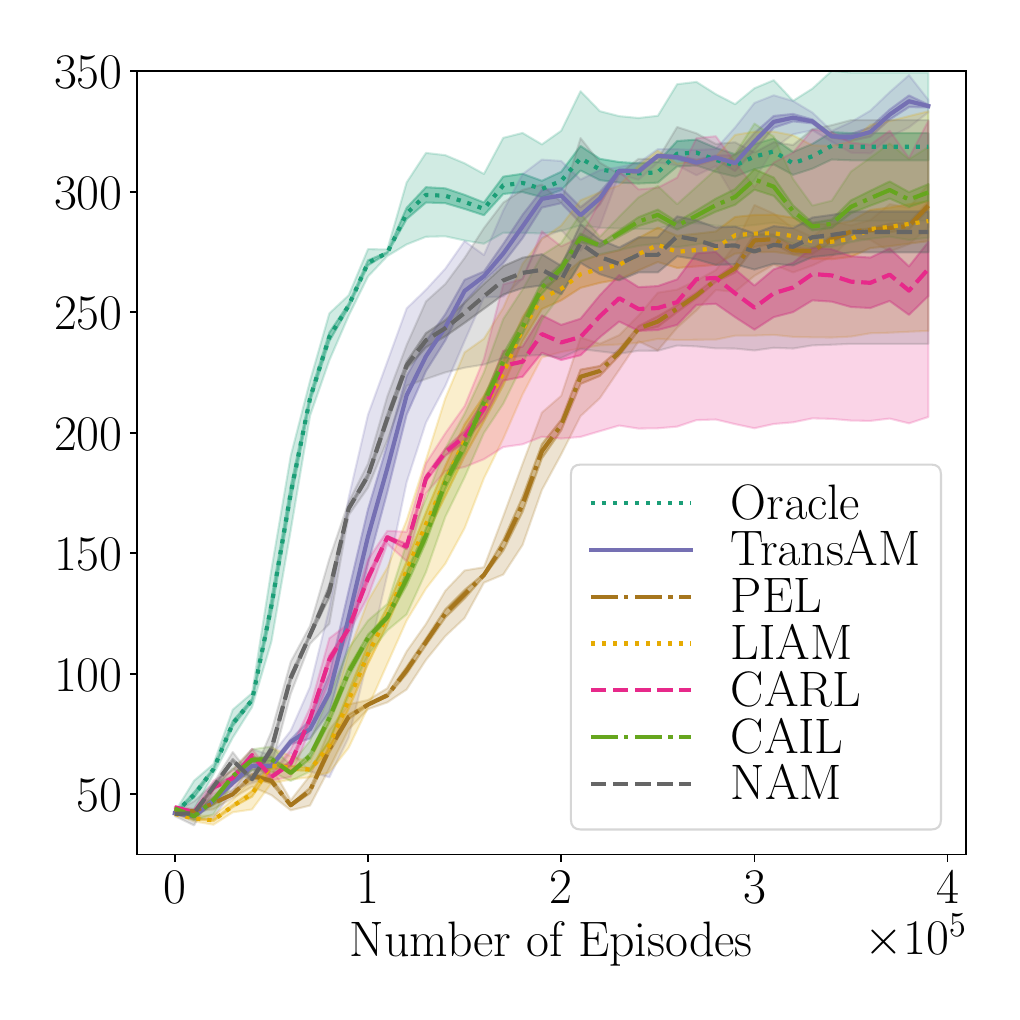}
        % \caption{Overcooked returns.}
        \label{fig:sub3}
    \end{subfigure}
    \begin{subfigure}[b]{.245\textwidth}
        \includegraphics[width=\linewidth]{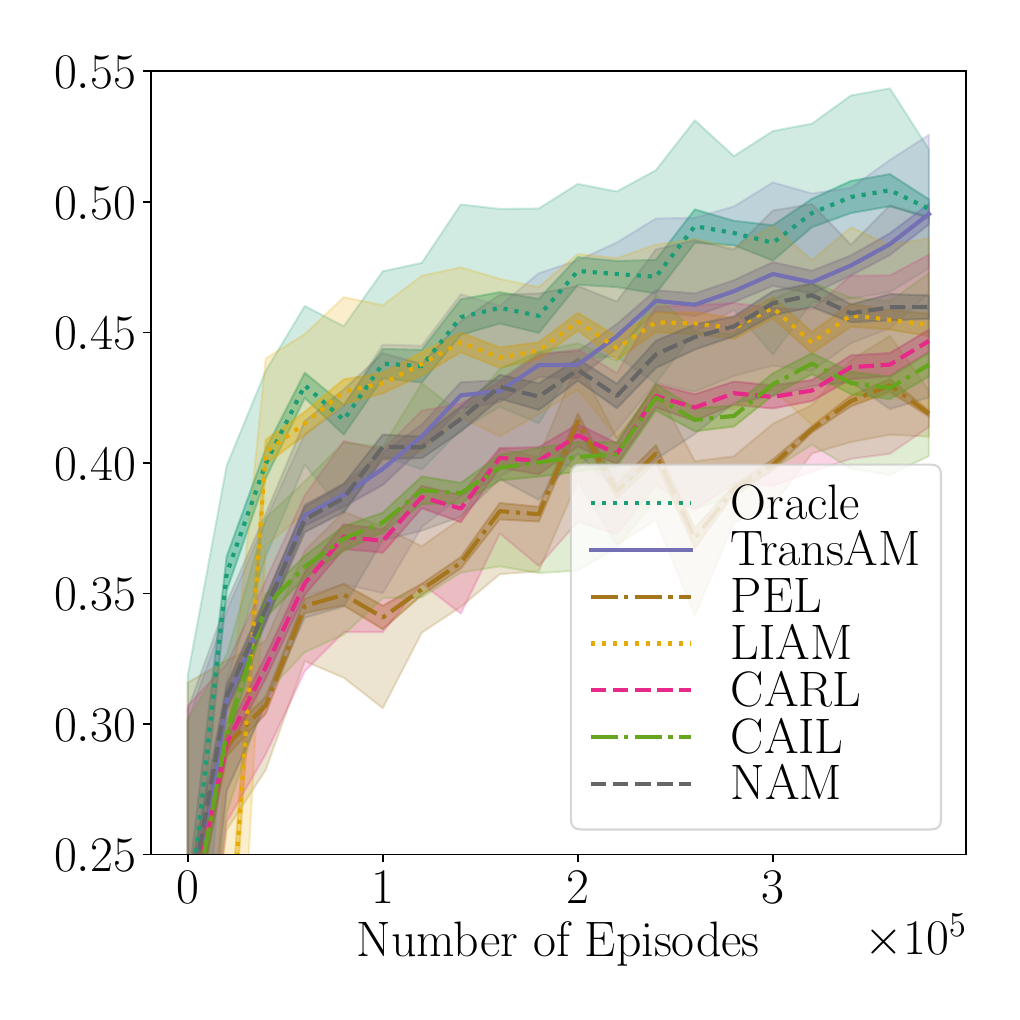}
        % \caption{LBF returns.}
        \label{fig:sub4}
    \end{subfigure}
    \begin{subfigure}[b]{.245\textwidth}
        \includegraphics[width=\linewidth]{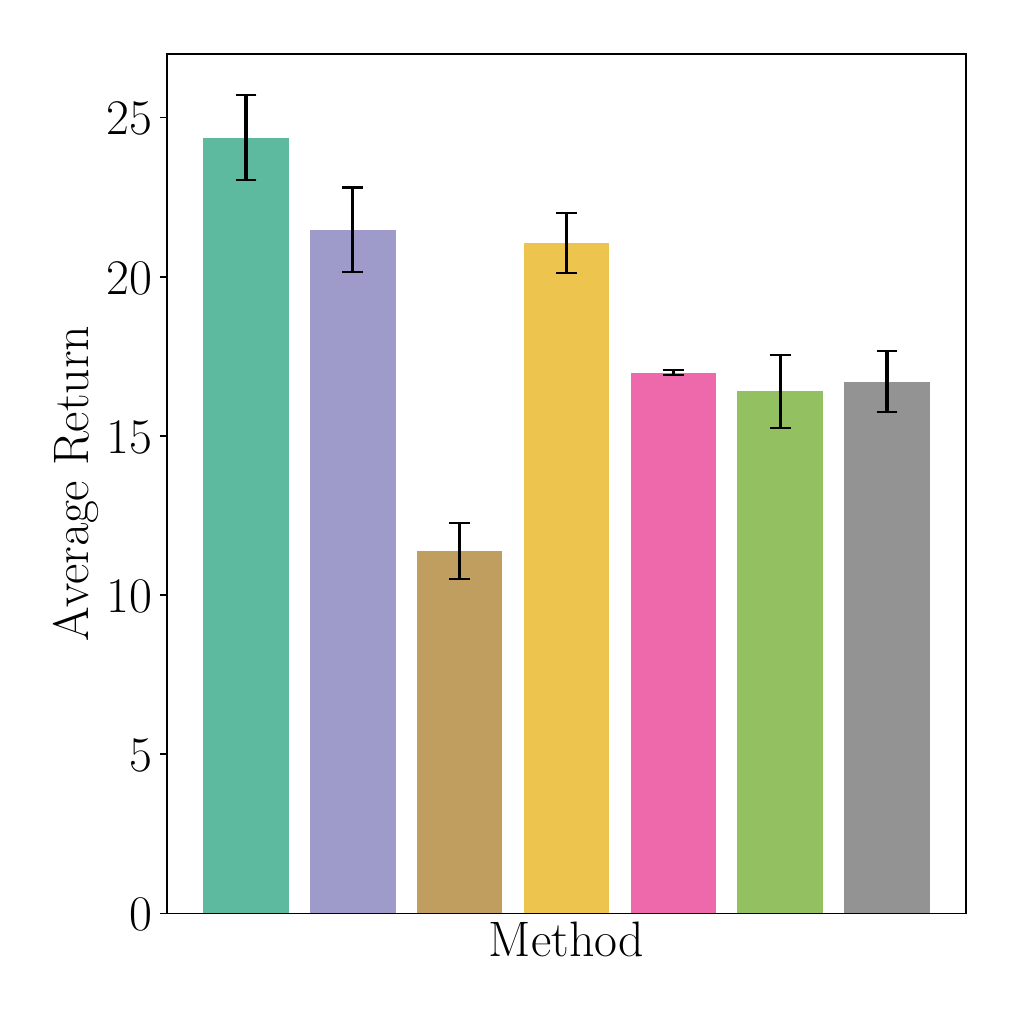}
        \caption{Tag returns.}
        \label{fig:sub1}
    \end{subfigure}
    \begin{subfigure}[b]{.245\textwidth}
        \includegraphics[width=\linewidth]{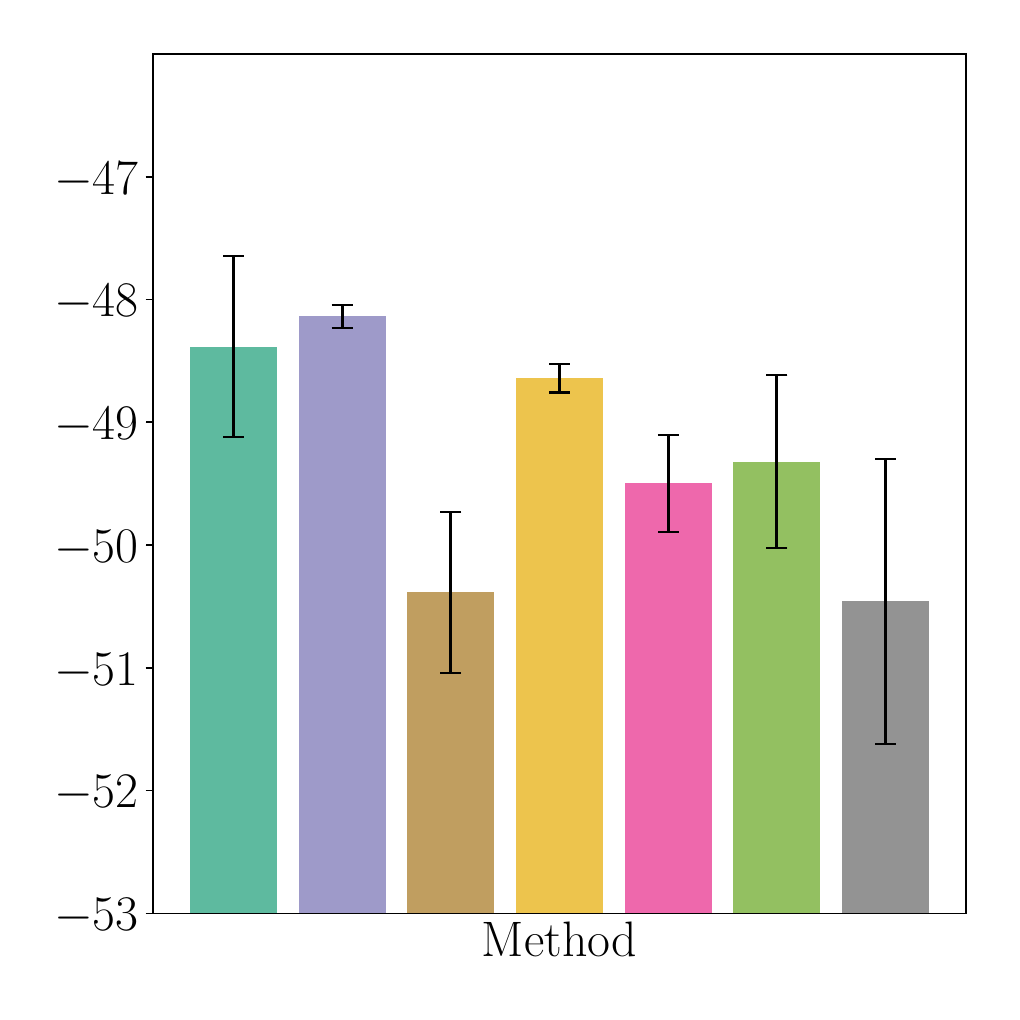}
        \caption{Spread returns.}
        \label{fig:sub2}
    \end{subfigure}
    \begin{subfigure}[b]{.245\textwidth}
        \includegraphics[width=\linewidth]{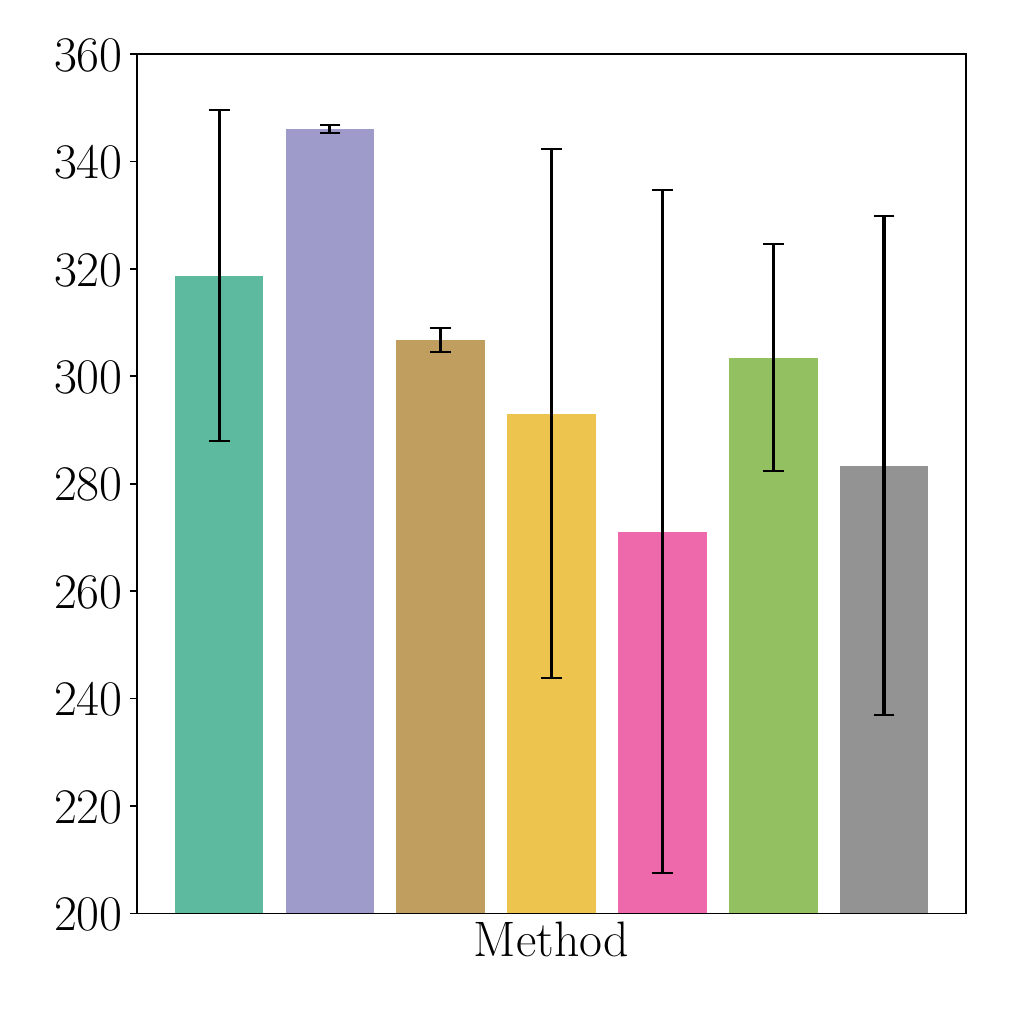}
        \caption{Overcooked returns.}
        \label{fig:sub3}
    \end{subfigure}
    \begin{subfigure}[b]{.245\textwidth}
        \includegraphics[width=\linewidth]{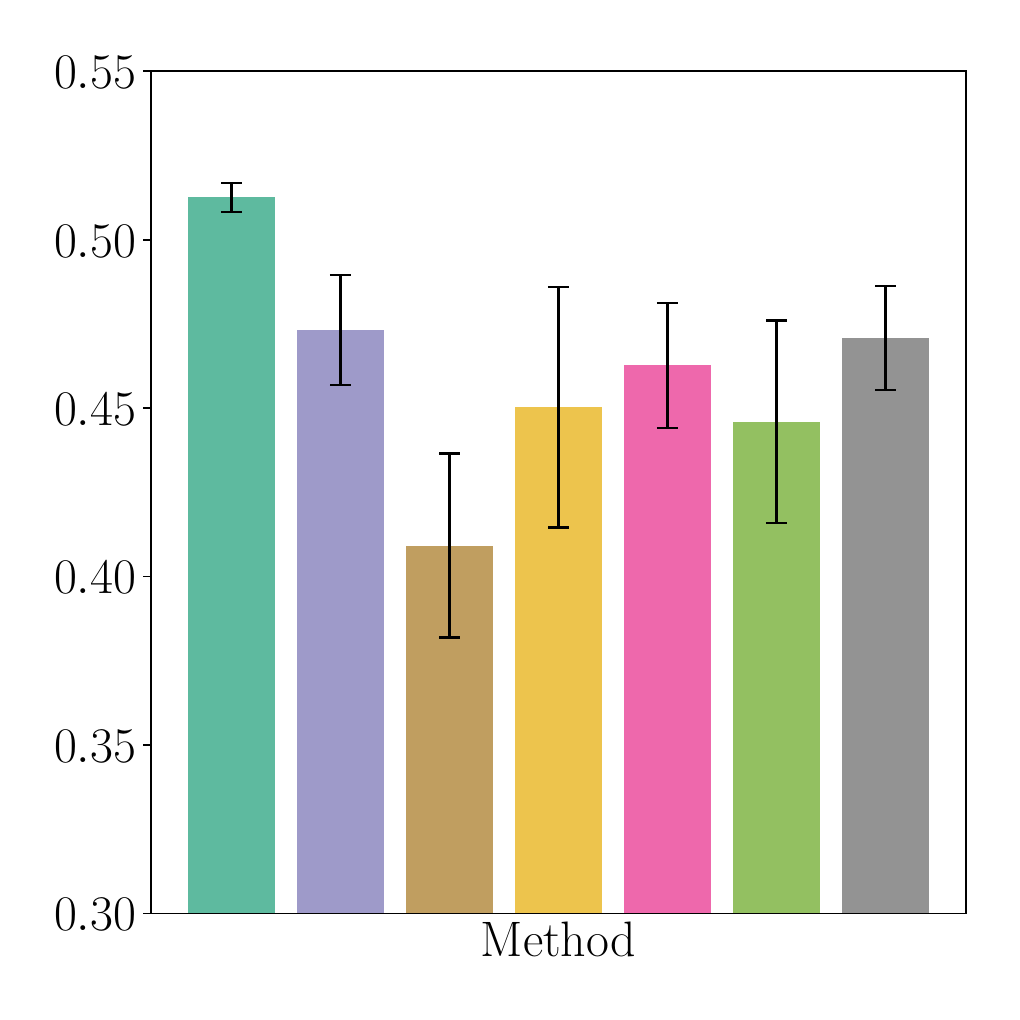}
        \caption{LBF returns.}
        \label{fig:sub4}
    \end{subfigure}
    \caption{\textbf{Average task returns.} (Top) Average episodic returns during training with 95\% confidence intervals across four experimental scenarios, evaluated over five random seeds. (Bottom) Mean and standard deviation of episodic returns over 100 evaluation episodes, also averaged across five random seeds.}
    \label{fig:returns}
\end{figure}

We compare our proposed method with several key baselines that represent a range of solutions in this space. Some baselines employ an explicit agent model, while others are implicit. These baselines can be categorized based on the amount of information available to the controlled agent about the other agents:

\begin{itemize}
    \item \textbf{No Agent Modeling (NAM):}\hspace{1ex} This baseline only has access to the controlled agent's current observation and last action.
    \item \textbf{Contrastive Agent Representation Learning (CARL):}\hspace{1ex} This baseline employs a recurrent encoder to embed the local information of the controlled agent into a vector space representing the joint policy. The encoder is trained using contrastive loss, specifically InfoNCE \citep{simclr}.
    \item \textbf{Conditional Agent Imitation Learning (CAIL):}\hspace{1ex} This baseline uses a recurrent backbone to embed local information into a vector space, which is then used to condition a policy imitation decoder.
    \item \textbf{Local Information Agent Modeling (LIAM):}\hspace{1ex} This baseline from \cite{liam} employs a recurrent encoder-decoder architecture to encode the controlled agent's local information into an embedding space. The decoder reconstructs other agents' observations and actions, but only the encoder is used during inference, restricting access to the controlled agent's information.
    \item \textbf{Policy Embedding Learning (PEL):}\hspace{1ex} Originally proposed in \cite{tao}, this approach uses a transformer-based architecture to encode an opponent's trajectory into a policy embedding space. It employs a generative loss for action reconstruction via conditional imitation learning and a contrastive InfoNCE loss to differentiate policies. We adapt this by encoding only the controlled agent's trajectory.
    \item \textbf{Oracle:}\hspace{1ex} This baseline assumes full access to other agents' trajectories, including observations and actions. The controlled agent conditions on a joint vector comprising its local observation, last action, and other agents' observations and actions. With no ambiguity in the intentions or strategies of the agents, this represents an upper performance baseline.
\end{itemize}

\subsection{Experimental Environments}
\subsubsection{Predator-Prey (Tag)}
We use a modified predator-prey environment from \citep{prey}, consisting of two large landmarks, three adversarial predator agents, and one controlled prey agent. The prey is faster, providing a strategic advantage. The prey receives a reward of \(+1\) if caught by a single adversary, while all adversaries receive \(-1\). If multiple adversaries capture the prey, the prey receives \(-1\) and the adversaries receive \(+1\). Additionally, the agent incurs a penalty \(-10\) for reaching the boundary.

\begin{figure}
    \centering
    \begin{subfigure}[b]{.245\textwidth}
        \includegraphics[width=\linewidth]{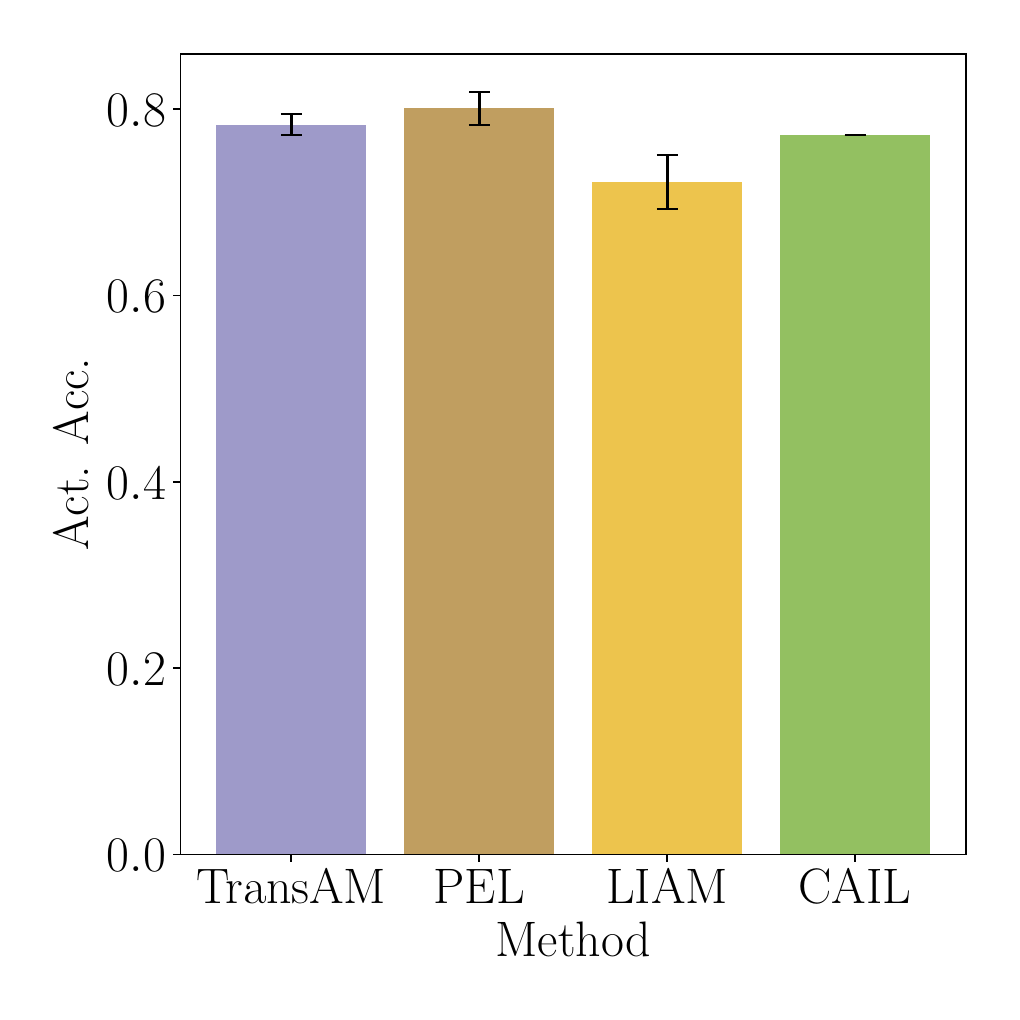}
        \caption{Tag.}
        \label{fig:sub1}
    \end{subfigure}
    \begin{subfigure}[b]{.245\textwidth}
        \includegraphics[width=\linewidth]{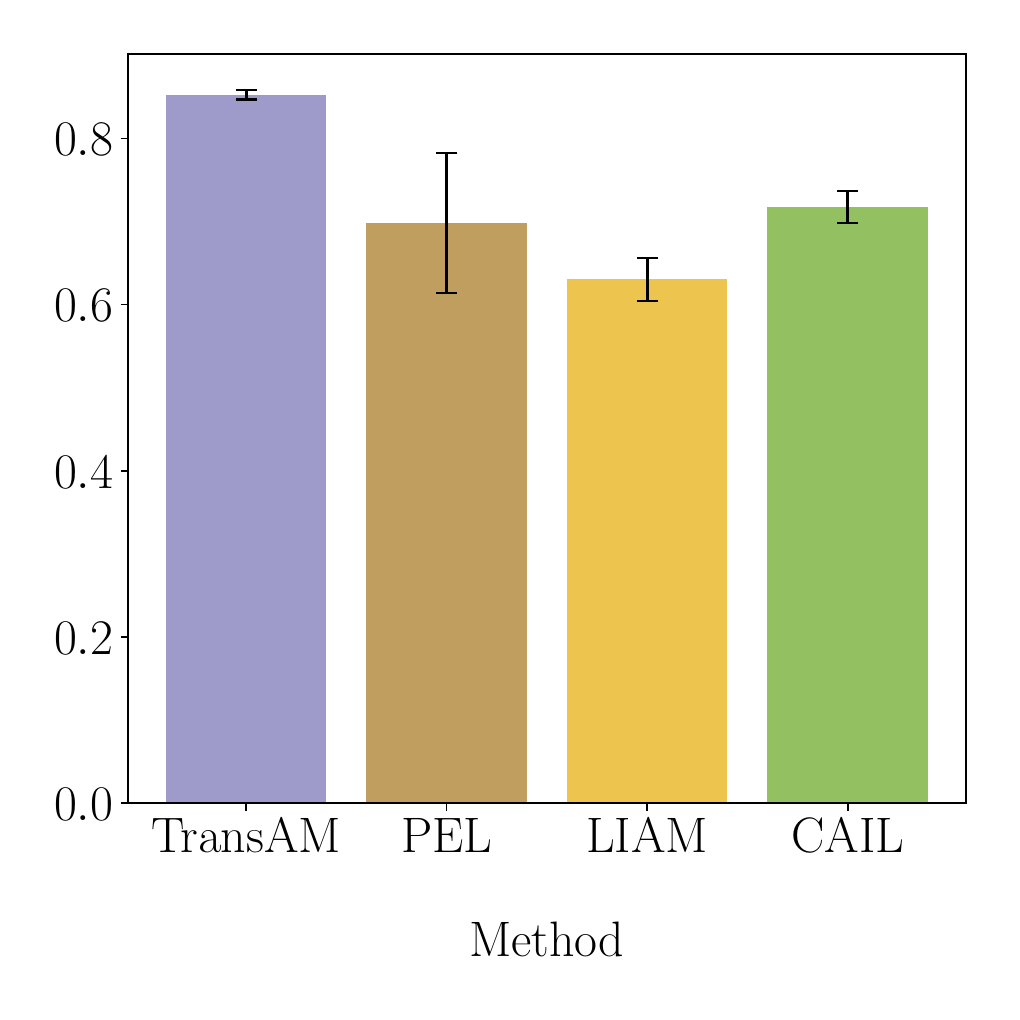}
        \caption{Spread.}
        \label{fig:sub2}
    \end{subfigure}
    \begin{subfigure}[b]{.245\textwidth}
        \includegraphics[width=\linewidth]{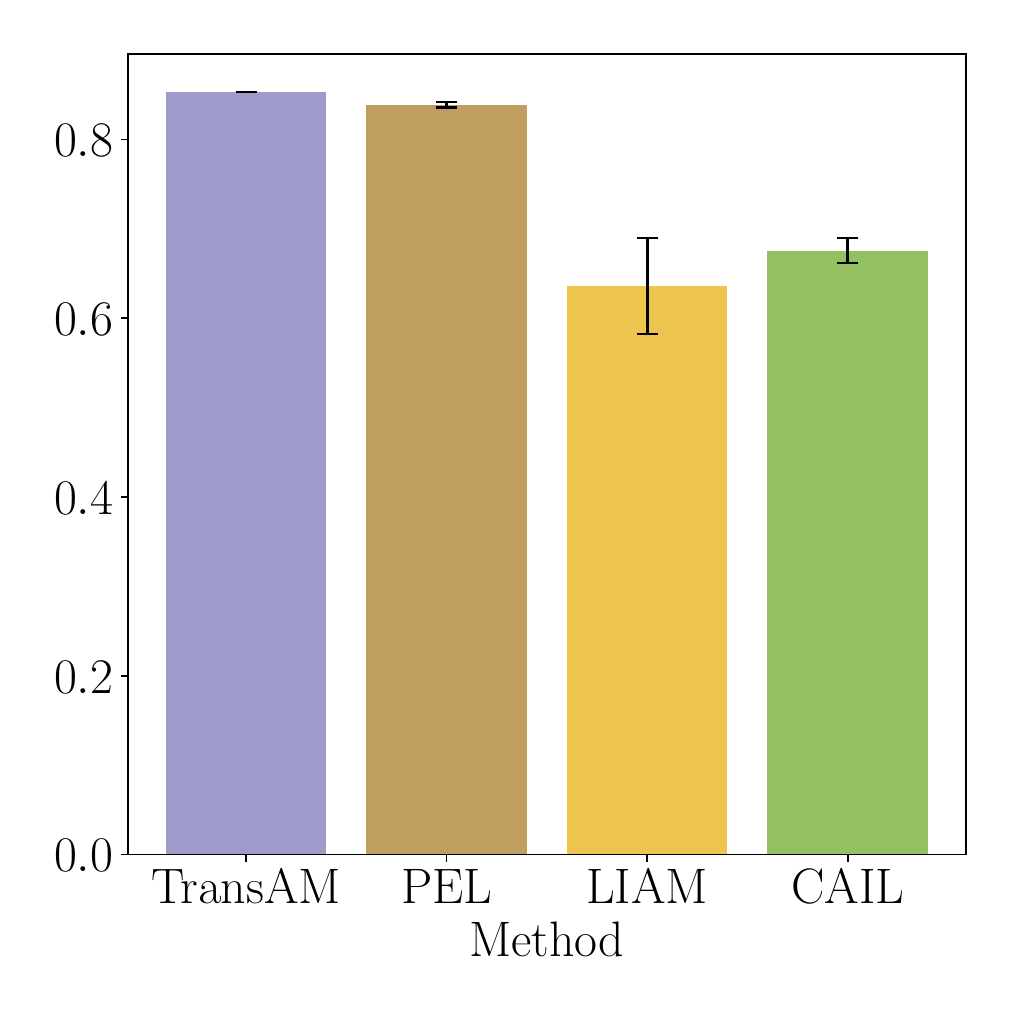}
        \caption{Overcooked.}
        \label{fig:sub3}
    \end{subfigure}
    \begin{subfigure}[b]{.245\textwidth}
        \includegraphics[width=\linewidth]{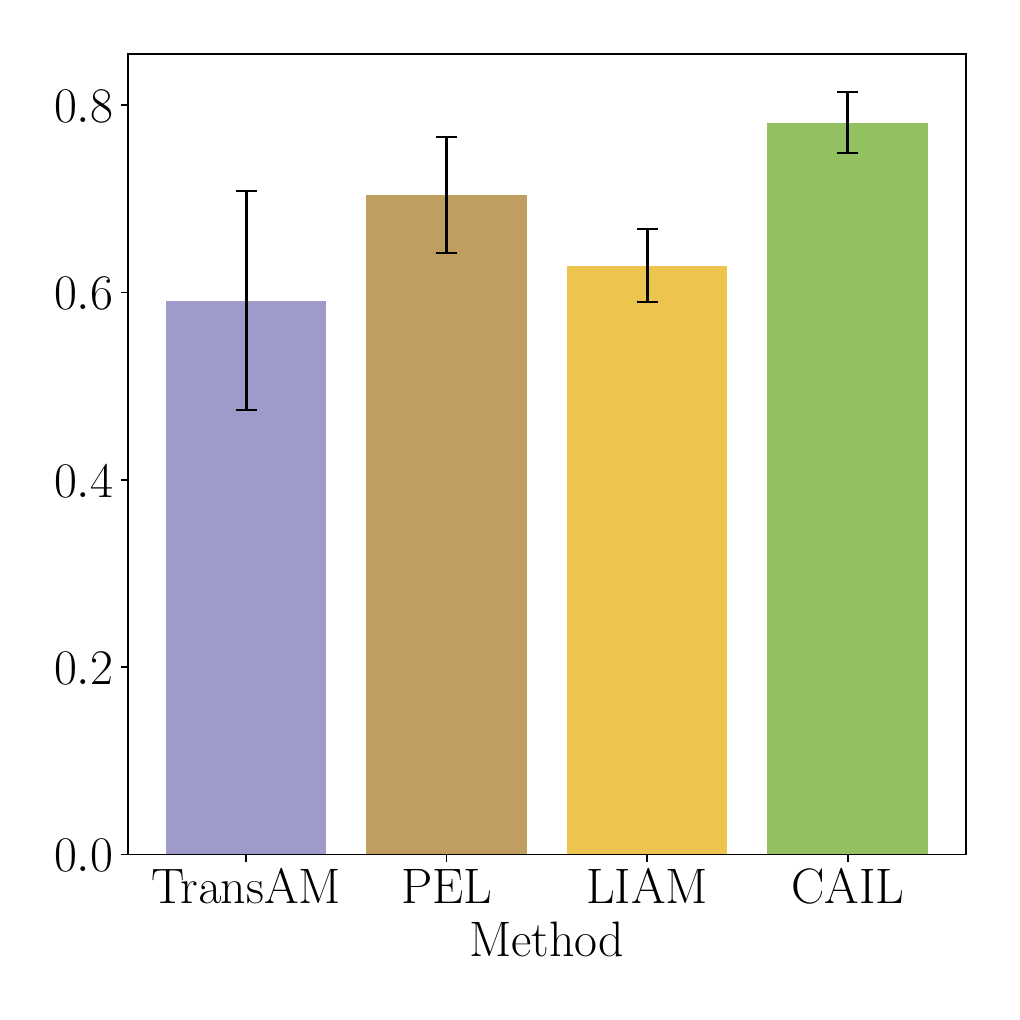}
        \caption{LBF.}
        \label{fig:sub4}
    \end{subfigure}
    \caption{\textbf{Agent action reconstruction accuracy.} We compute the mean and standard deviation of the other agents' action reconstruction accuracy for the relevant methods for all four environments averaged across five random seeds.}
    \label{fig:accuracies}
\end{figure}

\subsubsection{Cooperative Navigation (Spread)}
We use the original cooperative navigation scenario from \cite{mpe}, where three agents and three landmarks start from random positions. Agents must coordinate to cover all landmarks while avoiding collisions. The team’s reward is based on the sum of the minimum distances between agents and landmarks, with penalties for collisions.

\subsubsection{Overcooked}
We utilize the cramped room layout from the simplified Overcooked environment \citep{overcooked}, where two chefs collaborate in a confined kitchen to prepare and serve onion soup. The task requires executing a sequence of high-level actions, including placing onions in a pot (cooking for 20 timesteps), transferring soup to bowls, and serving. Each served soup grants both agents a reward of 20, with the objective of maximizing the number of soups served within 400 timesteps. Efficient coordination and multitasking are essential for optimal performance.

% For this task, we trained eight partner agents using PPO under different random seeds. We utilized reward shaping to induce different behaviors including specializing in dish pickup and placing onions in the pot, as well as no reward shaping to perform well across all subtasks.

% The average evaluation returns are shown in~\Cref{fig:return1b}. As expected, both NAM and Oracle perform consistently well. Similarly, CAIL struggles to outperform the lower baseline set by NAM. CARL is significantly better and achieves average returns towards the upper end of the baselines. LIAM performs slightly better yet again, nearly matching TransAM. This suggests that reconstructing both the joint observations and actions is key to obtaining good results. Finally, TransAM achieves an average return matching or exceeding the Oracle agent and once again converges quickly.

% The agent modeling results are depicted in~\Cref{fig:f1b}. Once again, TransAM achieves superior agent modeling metrics with CAIL slightly outperforming LIAM. Both CAIL and TransAM use cross-entropy loss to train the action reconstruction decoder, while LIAM uses negative log-likelihood. This is likely the reason for the slight dip in agent modeling performance from LIAM.

\subsubsection{Level-Based Foraging}
This scenario features a 20×20 gridworld with two agents and four food locations, each assigned a skill level. An agent can capture food if its skill level exceeds that of the food, and agents can also combine skill levels to capture higher-level food. This creates a mixed cooperative-competitive dynamic, where agents may collaborate for higher rewards or act independently for easier gains. Rewards are distributed based on each agent’s contribution to the total captured food. For instance, if one agent captures food of level 1 while the other captures levels 2, 3, and 4, their rewards are proportionally \( 1/(1+2+3+4) \) and \( (2+3+4)/(1+2+3+4) \), respectively.

% For this task, we train ten policies using MAPPO and IPPO using different random seeds. Additionally, we continuously reduce the discount factor to make some agents more myopic than others, which leads to agents who are incentivized to act greedily in favor of short-term rewards over long-term cooperation.

\begin{figure}
    \centering
    \begin{subfigure}[b]{.245\textwidth}
        \includegraphics[width=\linewidth]{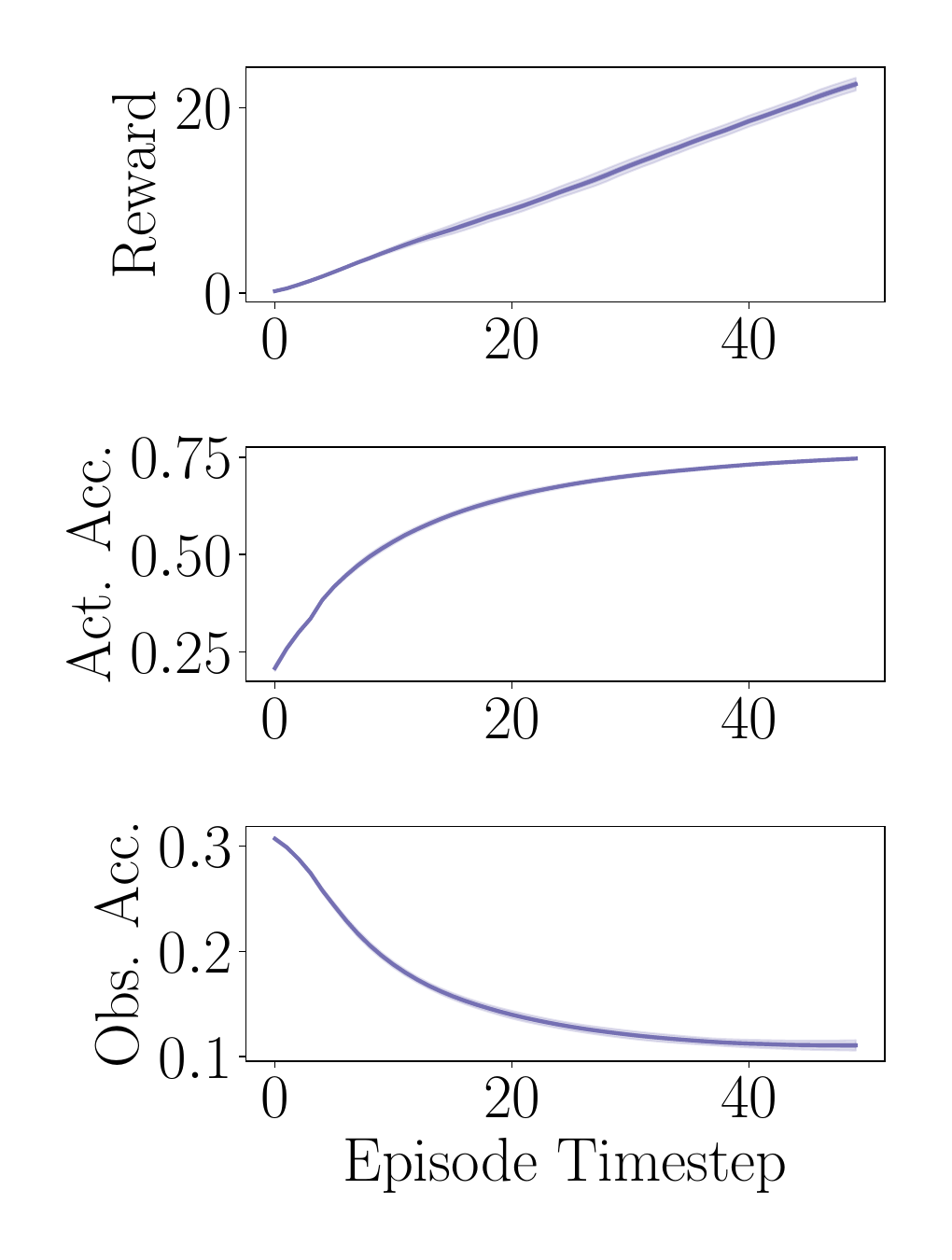}
        \caption{Tag returns.}
        \label{fig:sub1}
    \end{subfigure}
    \begin{subfigure}[b]{.245\textwidth}
        \includegraphics[width=\linewidth]{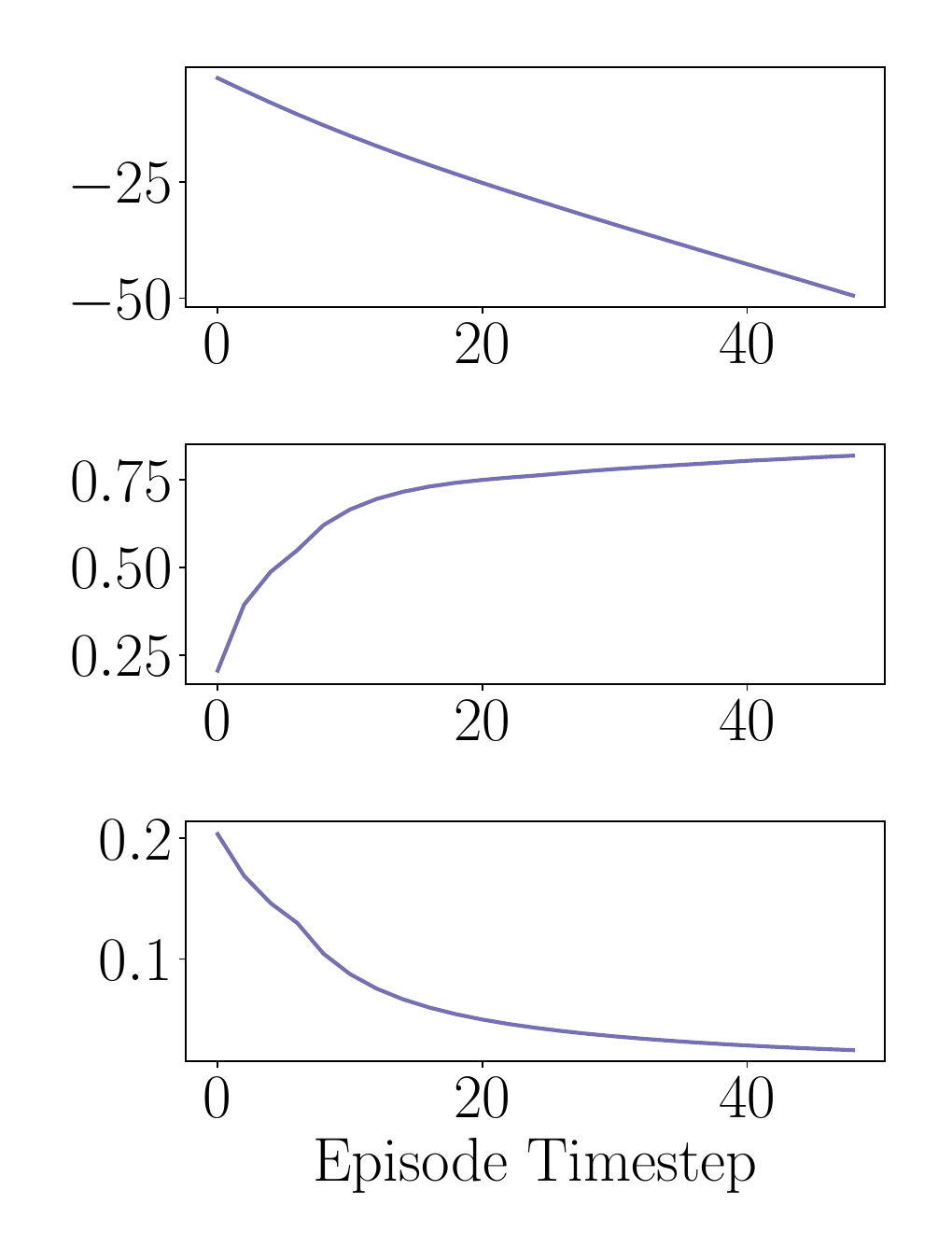}
        \caption{Spread returns.}
        \label{fig:sub2}
    \end{subfigure}
    \begin{subfigure}[b]{.245\textwidth}
        \includegraphics[width=\linewidth]{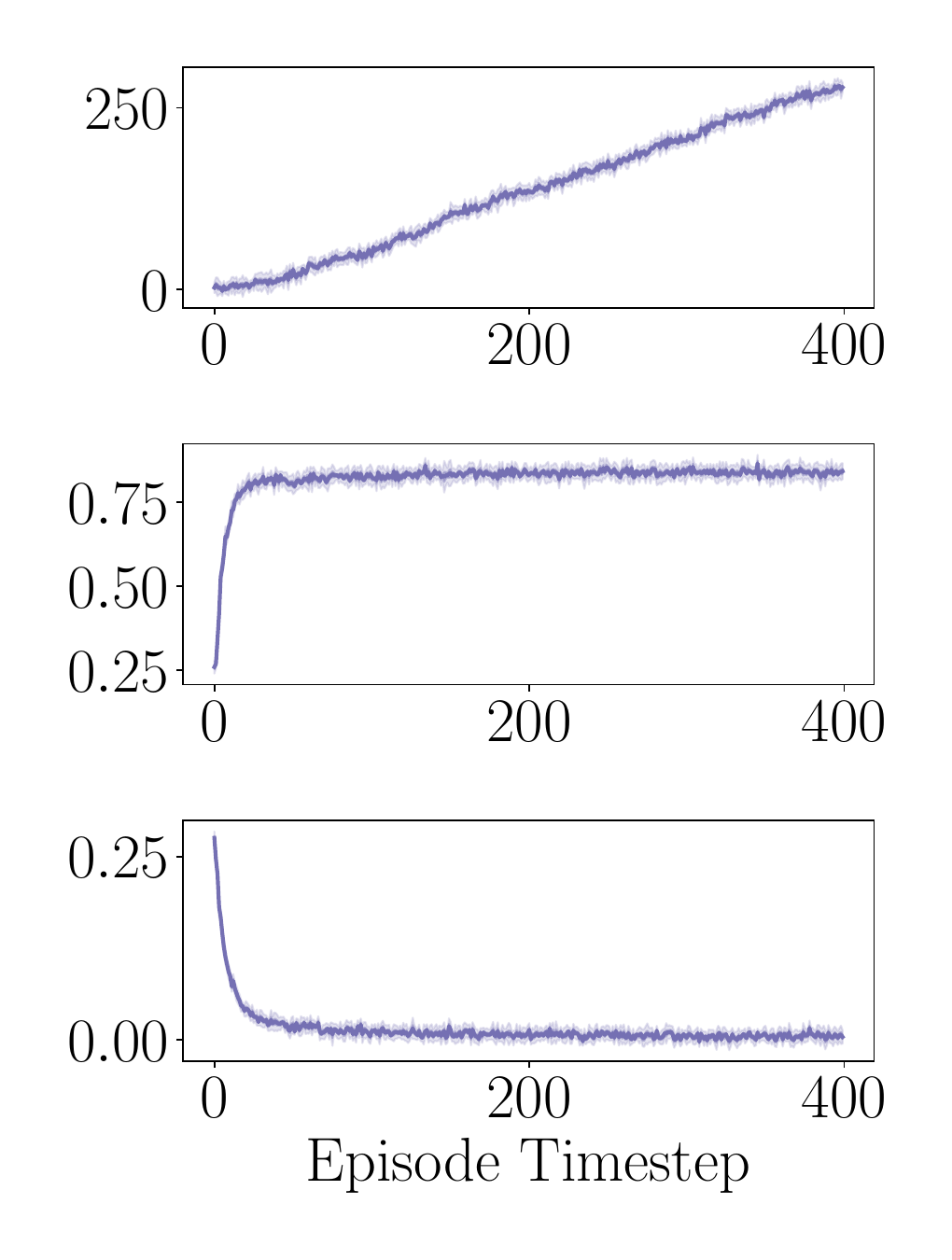}
        \caption{Overcooked returns.}
        \label{fig:sub3}
    \end{subfigure}
    \begin{subfigure}[b]{.245\textwidth}
        \includegraphics[width=\linewidth]{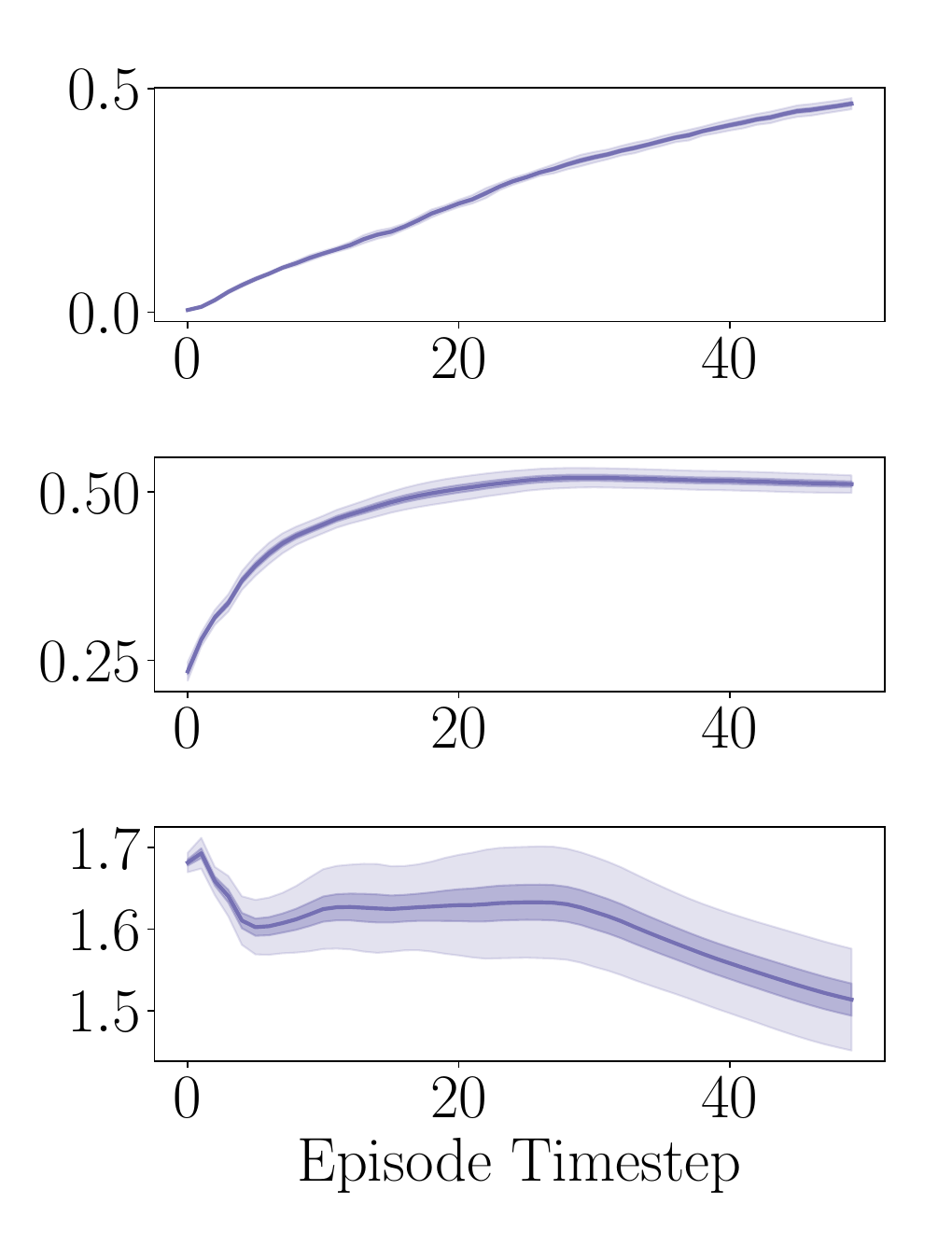}
        \caption{LBF returns.}
        \label{fig:sub4}
    \end{subfigure}
    \caption{\textbf{Evolution of \texttt{TransAM} performance across an episode.} We analyze the relationship between cumulative reward $\uparrow$ (top), agent action reconstruction accuracy $\uparrow$ (middle), and agent observation reconstruction accuracy as the mean-squared error $\downarrow$ (bottom) throughout an episode, averaged over 100 episodes. Note: The spread environment rewards are strictly negative, so the evolution of returns trends down. Error bars represent 95\% confidence intervals across 5 seeds.}
    \label{fig:episode}
\end{figure}

\subsection{Analysis}
\subsubsection{Task Returns}
% The average evaluation returns are shown in \ref{fig:returns}. As expected, both NAM and Oracle perform consistently well. Similarly, CAIL struggles to outperform the lower baseline set by NAM. CARL is significantly better and achieves average returns towards the upper end of the baselines. LIAM performs slightly better yet again, nearly matching TransAM. This suggests that reconstructing both the joint observations and actions is key to obtaining good results. Finally, TransAM achieves an average return matching or exceeding the Oracle agent and once again converges quickly.

Figure \ref{fig:returns} presents average evaluation returns. As expected, Oracle defines the upper performance bound. Notably, \texttt{TransAM} matches or exceeds Oracle across all environments, with LIAM performing similarly. Both benefit from encoding the controlled agent's actions and observations into more informative policy embeddings. NAM achieves moderate to low returns, likely due to the absence of auxiliary objectives. CAIL underperforms in predator-prey and level-based foraging but excels in cooperative navigation and Overcooked, highlighting the benefit of policy reconstruction in cooperative tasks. CARL performs moderately overall, particularly in competitive settings. PEL yields the lowest returns in most environments, suggesting that its combined generative and contrastive losses hinder final performance.

\subsubsection{Agent Modeling}
The agent modeling results for methods with action reconstruction capabilities are shown in Figure \ref{fig:accuracies}. \texttt{TransAM} consistently excels in reconstructing agent actions, outperforming all baselines in the two cooperative tasks, achieving competitive accuracy in the competitive task, but underperforming in the mixed setting. PEL matches or surpasses \texttt{TransAM} in three of four tasks, while CAIL performs comparably but struggles in cooperative environments. Both PEL and CAIL incorporate an imitation learning objective, with PEL additionally using a contrastive loss to better distinguish agent policies. However, this improved agent modeling performance comes at the cost of final task returns, suggesting a trade-off between policy reconstruction and maximizing the controlled agent’s reward. This trade-off is evident in LIAM, which lags behind other baselines in agent modeling but achieves significantly higher returns than PEL and CAIL. \texttt{TransAM} effectively balances both objectives, demonstrating competitive agent modeling while achieving the highest returns. Notably, \texttt{TransAM} is particularly well suited for strictly cooperative settings, where superior agent modeling performance strongly correlates with high returns, even surpassing the Oracle in some cases.

\subsubsection{Model Evaluation}
To understand the mechanisms behind the success of \texttt{TransAM}, we analyze its behavior throughout an episode in each test environment. Figure \ref{fig:episode} illustrates the relationship between the accuracy of the agent modeling and the cumulative reward. At the beginning of an episode, the model lacks context about the joint policy with which it is interacting, resulting in a policy embedding \( E^{o}_{t} \) that provides little additional information on the observation of the agent. However, as the episode progresses, the embeddings become more informative, improving agent modeling accuracy and leading to higher cumulative rewards.  

This relationship is further evident when comparing how quickly the model converges on other agents' trajectories to its performance relative to other baselines. For example, in the overcooked environment (Figure \ref{fig:episode} (c)), \texttt{TransAM} converges the fastest, aligning with its highest reward margin over the baselines (Figure \ref{fig:returns}(c)). In contrast, in the level-based foraging environment (Figure \ref{fig:episode}(d)), \texttt{TransAM} struggles to model agent behavior, which is correlated with its difficulty in outperforming other baselines (Figure \ref{fig:returns}(d)). These findings highlight the importance of designing adaptive agents that effectively model policies in environments with complex reward structures.

\begin{table*}
    \caption{\textbf{Model architecture ablation study results.} We test three variations of the model architecture on the cooperative navigation task and report the cumulative episodic return and the agent action reconstruction accuracy. The best results are shown in bold.}
    \begin{center}
        \begin{tabular}{lccc}
            \hline
            \multicolumn{1}{l}{\bf Method}  &\multicolumn{1}{l}{\bf Return}
            &\multicolumn{1}{l}{\bf Action Accuracy}
            \\ \hline
            TransAM & $\mathbf{-48.76}$ &  $\mathbf{85.72}$ \\
            TransAM-\textit{pool} & $-49.37$ & $61.67$ \\
            TransAM-\textit{fuse} & $-48.94$ & $78.68$ \\
            TransAM-\textit{im} & $-49.93$ & $72.08$ \\
            \hline
        \end{tabular}
    \end{center}
    %\caption{Sample table caption}
    \label{tab:exampleTable}
\end{table*}

\subsection{Model Architecture Ablation Study}

% We perform analysis on three ablated designs of \texttt{TransAM} in the cooperative navigation environment to test the primary components of the architecture; the multimodal embeddings, the aggregation of the embedding vectors, and the auxiliary training task. We consider the effect on cumulative episodic reward as well as agent action reconstruction accuracy. \texttt{TransAM}-\textit{fuse} concatenates the controlled agent's rewards, actions, and observations and learns a fused token embedding transformation compared to embedding tokens for each modality. \texttt{TransAM}-\textit{pool} uses an average pooling mechanism to merge all trajectory embeddings instead of using the most recent embedding. Finally, \texttt{TransAM}-\textit{im} uses conditional imitation learning as its decoder instead of predicting both agent observations and actions. The results for which are presented in Table \ref{tab:exampleTable}.

We analyze three ablated variants of \texttt{TransAM} in the cooperative navigation environment to evaluate the impact of its key architectural components: multimodal embeddings, embedding aggregation, and auxiliary training task. We assess their effects on cumulative episodic reward and agent action reconstruction accuracy.  
\begin{itemize}
    \item \texttt{TransAM}-\textit{fuse}: Concatenates the rewards, actions, and observations of the controlled agent into a single fused token embedding, rather than embedding the tokens separately for each modality.
    \item \texttt{TransAM}-\textit{pool}: Uses average pooling to merge all trajectory embeddings instead of relying on the most recent embedding.
    \item \texttt{TransAM}-\textit{im}: Employs conditional imitation learning as the decoder, predicting only agent actions rather than both observations and actions.
\end{itemize}
The results of this analysis are presented in Table \ref{tab:exampleTable}. First, we determine whether our local trajectory representation is beneficial by comparing it against \texttt{TransAM}-\textit{fuse}. This design achieves comparable returns; yet suffers in agent modeling tasks--indicating that separate token embeddings per modality are beneficial. Next, we consider the approach of pooling trajectory embeddings using \texttt{TransAM}-\textit{pool} as opposed to using the most recent embedding vectors to condition the controlled agent's policy. We observe that while this method incorporates information from the entire trajectory, it leads to poor performance for both episodic returns and action reconstruction accuracy, suggesting that recent transitions are more informative for identifying joint policies. Finally, we test whether the conditional imitation learning decoder in \texttt{TransAM}-\textit{im} provides a benefit over decoding both the observations and actions of the agent. This produces the worst returns and second-lowest modeling accuracy, reinforcing the importance of reconstructing both observations and actions. 
% This is confirmed by the fact that LIAM and \texttt{TransAM} consistently achieve top-average episodic returns.
\section{Conclusion and Future Work}
\label{sec:conclusion}

In this paper, we introduced \texttt{TransAM}, a transformer-based agent modeling architecture that operates without access to other agents' information at execution time, ensuring full decentralization of the controlled agent. Using a transformer, \texttt{TransAM} effectively extracts and utilizes features from the controlled agent's episodic trajectory. We demonstrated its effectiveness across multiple environments, including Predator-Prey and Cooperative Navigation from the multi-agent particle environments, as well as Overcooked and Level-Based Foraging.  

For future work, we aim to investigate the scalability of agent modeling techniques in larger multi-agent systems. Additionally, we seek to explore recursive reasoning domains, where agents must model others while accounting for the fact that their opponents are also performing agent modeling.

\subsubsection*{Acknowledgments}
\label{sec:ack}
This work was supported by the Office of Naval Research under Grant N000142412405 and the Army Research Office under Grants W911NF2110103 and W911NF2310363.

%%%%%%%%%%%%%%%%%%%%%%%%%%%%%%%%%%%%%%%%%%%%%%%%%%%%%%%%%%%%%%%%
%% Appendices
%%%%%%%%%%%%%%%%%%%%%%%%%%%%%%%%%%%%%%%%%%%%%%%%%%%%%%%%%%%%%%%%
% \appendix

% \section{The first appendix}
% \label{sec:appendix1}
% This is an example of an appendix. 

% \noindent \textbf{Note:} Appendices appear before the references and are viewed as part of the ``main text'' and are subject to the 8--12 page limit, are peer reviewed, and can contain content central to the claims of the paper. 

% \section{The second appendix}
% \label{sec:appendix2}
% This is an example of a second appendix. If there is only a single section in the appendix, you may simply call it ``Appendix'' as follows:

% \section*{Appendix}
% % No label, since this can't be referenced meaningfully with \ref{}.
% This format should only be used if there is a single appendix (unlike in this document).

% \label{sec:ack}
% Use unnumbered third level headings for the acknowledgments. All acknowledgments, including those to funding agencies, go at the end of the paper. Only add this information once your submission is accepted and deanonymized. The acknowledgments do not count towards the 8--12 page limit.

%%%%%%%%%%%%%%%%%%%%%%%%%%%%%%%%%%%%%%%%%%%%%%%%%%%%%%%%%%%%%%%%
%% NOTE: THIS MARKS THE END OF THE "MAIN TEXT"
%%%%%%%%%%%%%%%%%%%%%%%%%%%%%%%%%%%%%%%%%%%%%%%%%%%%%%%%%%%%%%%%

%%%%%%%%%%%%%%%%%%%%%%%%%%%%%%%%%%%%%%%%%%%%%%%%%%%%%%%%%%%%%%%%
%% Bibliography
%%%%%%%%%%%%%%%%%%%%%%%%%%%%%%%%%%%%%%%%%%%%%%%%%%%%%%%%%%%%%%%%
\bibliography{main}
\bibliographystyle{rlj}

\end{document}